\documentclass[11pt]{article}
\usepackage{graphicx}
\usepackage{hep}
\usepackage{amsmath}
\usepackage{amssymb}
\usepackage{cite,color,url}
\usepackage[colorlinks=true
,urlcolor=blue
,anchorcolor=blue
,citecolor=blue
,filecolor=blue
,linkcolor=blue
,menucolor=blue
,linktocpage=true
,pdfproducer=medialab
,pdfa=true
]{hyperref}
\usepackage{a4wide}
\usepackage{rotating}
\usepackage{multirow}
\usepackage{verbatim}

\newcommand{\modmin}{\ensuremath{mod^{-}}}
\newcommand{\pMSSM}{\ensuremath{pMSSM^{c}}}

\begin{document}

\def\thefootnote{\fnsymbol{footnote}}

\def\Title#1{\begin{center} {\Large\bf #1 } \end{center}}
\def\Author#1{\begin{center}{ \sc #1} \end{center}}
\def\Address#1{\begin{center}{ \it #1} \end{center}}
\def\andauth{\begin{center}{and} \end{center}}
\def\submit#1{\begin{center}Submitted to {\sl #1} \end{center}}
\newcommand\pubblock{\rightline{\begin{tabular}{l} \pubnumber\\
        \pubdate\\ \hepnumber \end{tabular}}}

\Title{Top-quark Polarization and Asymmetries at the LHC\\[.5em]
in the Effective Description of Squark Interactions}

\vspace{1cm}

\Author{
A. Abrahantes%
\footnote{\tt \href{mailto:arian@unizar.es}{arian@unizar.es}}%
, E. Arganda%
\footnote{\tt \href{mailto:ernesto.arganda@unizar.es}{ernesto.arganda@unizar.es}}%
, S. Pe\~{n}aranda%
\footnote{\tt \href{mailto:siannah@unizar.es}{siannah@unizar.es}}%
}
\Address{\textsl{
Departamento de F{\'\i}sica Te{\'o}rica,
Facultad de Ciencias,\\ Universidad de Zaragoza,
            E-50009 Zaragoza, Spain}}

\vspace*{0.1cm}

\begin{abstract}
\noindent 
A detailed study of top-quark polarizations and $t \bar t$ charge
asymmetries, induced by top-squark-pair production at the LHC
and the subsequent decays $\tilde t \to t \tilde \chi_1^0$, is performed 
within the effective description of squark interactions,
which includes the effective Yukawa couplings and another logarithmic
term encoding the supersymmetry breaking. This effective approach is 
more suitable for its introduction into Monte-Carlo simulations and 
we make use of its implementation in {\tt MadGraph} in order to
investigate the possibilities of the charge asymmetry $A_\text{C}$, 
measured at the LHC and consistent with SM expectations,
to discriminate among different SUSY scenarios and analyze the 
implications of these scenarios in the top polarizations and related observables.
\end{abstract}

\def\thefootnote{\arabic{footnote}}
\setcounter{page}{0}
\setcounter{footnote}{0}

\newpage

\section{Introduction}
\label{intro}

The Standard Model (SM) of the electroweak and strong interactions is the present paradigm of particle physics and provides a very good description of all data collected so far at
hadron and lepton colliders~\cite{Beringer:1900zz}, including the recent discovery of the 
SM-like Higgs boson at the LHC~\cite{Aad:2012tfa,Chatrchyan:2012ufa}. 
However, there are arguments against the SM being the fundamental model of particle interactions~\cite{Haber:1993kz}, giving rise to the investigation of competing extended models. Among the alternatives of physics beyond the SM, 
one of the most promising possibilities is supersymmetry 
(SUSY)~\cite{Nilles:1983ge,Haber:1984rc,Lahanas:1986uc,Ferrara:1987ju}, 
which leads to a renormalizable field theory with precisely calculable predictions 
to be tested in present and future experiments and 
whose simplest realization at the electroweak scale is the Minimal Supersymmetric 
Standard Model (MSSM)~\cite{Chung:2003fi}. The MSSM predicts the existence of superpartners 
for each SM particle: squarks/sleptons, 
gauginos and higgsinos are the partners of quarks/leptons, gauge and higgs bosons, 
respectively. 
The Higgs sector contains two scalars doublets, with a spectrum that includes three neutral Higgs bosons ($h$, $H$, $A$) 
and one charged Higgs pair ($H^\pm$)~\cite{Carena:2002es}, with the lightest MSSM 
Higgs boson $h$ being completely 
consistent with the discovered SM-like Higgs boson with mass 
$m_{h_\text{SM}} = 125.9$ GeV (see e.g.~\cite{Hahn:2013ria}).

In this work we focus on the properties of the top-squarks, the SUSY partners of SM top quarks. 
In particular, we concentrate on the top-squark decay channels involving neutralinos, the fermionic neutral superpartners 
of the electroweak gauge and Higgs bosons. Once produced, top-squarks will decay in a way dependent on the model 
parameters (see e.g.~\cite{Bartl:1994bu}). If the decay channels into gluinos and quarks or into other squarks and 
gauge or Higgs bosons are not kinematically allowed, the main decay channels of top-squarks are their partial decays into 
charginos and $b$ quarks ($\tilde t \to b \tilde \chi^\pm$) or into neutralinos and top quarks ($\tilde t \to t \tilde \chi^0$). 
Some of those channels are expected to be always open, given the large mass difference between quarks and top-squarks, 
and that the charginos/neutralinos are expected to be lighter than the top-squarks in the majority of SUSY-breaking models. 
In the few cases in which these channels are closed, the top-squarks will decay through flavor-changing 
neutral channels~\cite{Hikasa:1987db,Han:2003qe,delAguila:2008iz}, or through three- or four-body decay channels involving a 
non-resonant SUSY particle~\cite{Porod:1996at,Porod:1998yp,Boehm:1999tr,Djouadi:2000aq,Das:2001kd,Djouadi:2000bx}. 
Here we will concentrate on the top-squark decay channels involving neutralinos and top quarks, within the effective description 
developed in~\cite{Guasch:2008fs}, more suitable for their introduction in the Monte-Carlo programs used in experimental 
analysis. This computation combines the effective description (which includes higher order terms) with the complete 
one-loop description (which includes all kinetic and mass-effects factors) and defines a new effective coupling. 
It includes a non-decoupling logarithmic gluino mass term, which implies a deviation of the higgsino/gaugino and 
Higgs/gauge couplings equality predicted by exact SUSY. This deviation is important and has to be taken into account in 
the experimental measurement of SUSY relations. It is showed in~\cite{Guasch:2008fs} that the effective description 
approximates the improved description within a 10$\%$ precision, except in special uninteresting corners of the 
parameter space, where the corresponding branching ratios are practically zero. Whilst the results in~\cite{Guasch:2008fs} 
apply the description only to squark decays, a more recent work~\cite{Abrahantes:2012xm} expands the 
results of~\cite{Guasch:2008fs} by applying those results to the production cross section of squarks at the LHC. 
In~\cite{Abrahantes:2012xm,sqeffective}, this effective description has been implemented in {\tt MadGraph}~\cite{Alwall:2007st,Alwall:2011uj} 
MSSM framework~\cite{Cho:2006sx}, in order to be applied to the partial decay widths of squarks into charginos and 
neutralinos and compute the corresponding cross sections.

After the discovery of the top-quark~\cite{Abe:1995hr,Abachi:1995iq}, 
top-quark physics has entered the era of precision
measurements. Among the many measurements performed, the $t \bar t$ forward-backward 
asymmetry has received an special 
attention because of a disagreement with respect to the SM 
predictions~\cite{Kuhn:1998jr, Kuhn:1998kw,Kuhn:1998pn,Rodrigo:1998gd,Antunano:2007da,Kuhn:2011ri,Ahrens:2011uf,Hollik:2011ps}.
The $t \bar t $ lepton based asymmetries are above the SM 
as well~\cite{Aaltonen:2013vaf,Abazov:2013wxa,Abazov:2014oea,Aaltonen:2014eva}.
These discrepancies have motivated a plethora of new physics explanations.
However, most of the present precision $t \bar t$ measurements at the LHC 
exclude some of the simplest forms of the
new physics models proposed. In this work we analyze some effects of SUSY in new
physics $t \bar t$ observables.

An important aspect of the top-squark phenomenology is the possible contribution 
of the decay channel 
$\tilde t \to t \tilde \chi_1^0$ to new physics $t \bar t$ observables at the 
LHC~\cite{Boos:2003vf,Perelstein:2008zt,Bhattacherjee:2012ir,Belanger:2012tm}, 
as charge asymmetries or top-quark polarization. 
The presence of top-quarks from this decay could mean non-zero polarizations 
in the resulting final state at the LHC. 
Since the top-quark decay occurs before hadronization, the polarization can 
have important implications for the kinematic
 distributions of the final particles, and hence on the search strategies 
for the top-squarks~\cite{Belanger:2012tm}. 
 The longitudinal polarization of top-quarks coming from top-squark decays 
into neutralinos depends on the mass difference 
 between the top-squark and the neutralino, as well as on the mixing in both 
sectors, and can take any value between $-1$ and 
 $+1$, while the measurements of ATLAS~\cite{Aad:2013ksa} and CMS~\cite{Chatrchyan:2013wua} are in good agreement with the SM prediction of negligible top-quark polarization.
 Therefore, polarization studies may supply information about different SUSY 
scenarios~\cite{Belanger:2013gha}.
 On the other hand, explanations of the top-quark 
 forward-backward asymmetry $A_\text{FB}$ observed at the 
 Tevatron~\cite{Aaltonen:2012it,CDF:2013gna,Aaltonen:2013vaf,Abazov:2013wxa,Abazov:2014cca},
 which exceeds the SM predictions~\cite{Kuhn:1998pn,Rodrigo:1998gd, Kuhn:1998kw,Kuhn:1998jr, Ahrens:2011uf,Hollik:2011ps,Kuhn:2011ri,Bernreuther:2012sx}, 
 must take into account 
 the measurements of the $t \bar t$ charge asymmetry $A_\text{C}$ at the 
 LHC~\cite{ATLAS:2012sla,Aad:2013cea,Chatrchyan:2014yta} which are consistent with the SM expectations
 and tightly correlated with $A_\text{FB}$~\cite{AguilarSaavedra:2011hz,AguilarSaavedra:2011ug,Alvarez:2011hi,AguilarSaavedra:2011ck,AguilarSaavedra:2012va,AguilarSaavedra:2012rx,Aguilar-Saavedra:2013rza,Baumgart:2013yra,Aguilar-Saavedra:2014yea},
 given that both $t \bar t$ observables could receive contributions from 
top-squark pair production decaying into neutralinos and top-quarks.
 However, it is possible to have an excess in $A_\text{FB}$ and no excess in $A_\text{C}$
 if some cancellation occurs~\cite{AguilarSaavedra:2012va,Drobnak:2012cz,Leskow:2013wpa,Alvarez:2012ca,Drobnak:2012rb}
 or even exhibit differences in $t \bar t \gamma$ 
production~\cite{AguilarSaavedra:2011cp,Aguilar-Saavedra:2014vta}.
 We will not try in this work to search for particular SUSY scenarios which 
 would fit better the asymmetry $A_\text{FB}$ measured at the Tevatron and analyze their $A_\text{C}$ 
 predictions at the LHC. Our purpose here will be rather to look into the
   consistency of the proposed SUSY scenarios with the SM expectations for the charge asymmetries 
 defined in~\cite{Kuhn:2011ri} and then study the implications 
 of these scenarios on the top-quark polarizations and related observables at the LHC,
 which may help to discriminate among them and the SM. We put special emphasis on the way the effective
  description of squarks interactions affects these observables.

The paper is organized as follows. In section~\ref{th-frame} we review 
the most relevant features
of the effective description of squark/chargino/neutralino interactions.
The new physics $t \bar t$ observables related to the top-quark polarizations and
charge asymmetries at the LHC are presented in section~\ref{observables}. 
Section~\ref{num-results}
is devoted to the numerical analysis of $t \bar t$ charge asymmetries
and top-quark polarization observables
for several SUSY scenarios, comparing these results to the SM
predictions calculated at next to leading order (NLO).
In the end, the conclusions and final comments are summarized in section~\ref{conclusions}.

\section{Theoretical framework: effective description approximation}
\label{th-frame}

Here we work within the MSSM framework. It is well known that QCD corrections
to the squark partial decay widths into charginos and
neutralinos can be numerically large, specially in
certain regions of the parameter space~\cite{Kraml:1996kz,Djouadi:1996wt,Guasch:1999tk,Guasch:2002ez}. 
The complete one-loop corrections to squark partial decay
  widths are already available~\cite{Guasch:2002ez,SFdecay}, but their
complicated expressions are not suitable for the introduction in
Monte-Carlo programs used for experimental analysis. 
An effective description of squark/chargino/neutralino couplings, 
simple to write and to introduce in computer codes, 
was given in~\cite{Guasch:2008fs}. This description contains the 
large one-loop corrections from the
finite threshold corrections to the quark masses, but it also contains
higher order corrections including another logarithmic term which encodes 
the supersymmetry breaking.  
The above effective description have been 
recently implemented in {\tt MadGraph} MSSM framework~\cite{Abrahantes:2012xm}. 
In this article we use this implementation to discuss the effects
of the radiative corrections included in the effective description of squark interactions on
top polarization and top-quark asymmetries. 

In the following we present the crucial expressions of the effective approach which
have been included in {\tt MadGraph}
package~\cite{Alwall:2007st,Alwall:2011uj}. We briefly introduce the tree-level Lagrangian of
  interactions of the quark-squark-chargino/neutralino and then, an
  extract of the analysis in~\cite{Guasch:2008fs} depicting the
  effective description of the squarks interactions is presented.
 The tree-level interaction Lagrangian between fermion-sfermion-(chargino or neutralino)
reads~\cite{Guasch:2002ez}
    \begin{eqnarray}
      \label{eq:Lqsqcn}
      {\cal L}_{\chi \sfr f'}&=&\sum_{a=1,2}\sum_{r} {\cal L}_{\chi_r
      \sfr_a f'} + \mbox{ h.c.}\,\,,\nonumber\\
     {\cal L}_{\chi_r \sfr_a f'}&=& -g\,\sfr_a^* \bar{\chi}_r
      \left(A_{+ar}^{(f)}\pl +  A_{-ar}^{(f)}\pr\right) f'\,\,.
    \end{eqnarray}
Here we have adopted a compact notation, where $f'$ is either $f$ or its 
$SU(2)_L$ partner for $\chi_r$ being a neutralino or a chargino,
respectively. Roman characters $a,b\ldots$ are reserved for sfermion
indices and $i,j,\ldots$ for chargino indices, Greek indices
$\alpha,\beta,\ldots$ denote neutralinos, Roman indices $r,s\ldots$ indicate either a chargino or a
   neutralino. For example, the top-squark interactions with charginos
   are obtained by replacing $f\to t$, $f'\to b$, $\chi_r\to \cmin_r$,
   $r=1,2$. The coupling matrices that encode the dynamics are given by:
\begin{eqnarray}
        \label{eq:V1Apm}
        \Apit &=& \Rot\Vo^*-\lt\Rtt\Vt^*\, ,\nonumber\\
        \Amit &=& -\lb\Rot\Ut\, ,\nonumber\\
        \Apat &=&\frac{1}{\sqrt{2}} \left(
            \Rot\left(\Nt^*+\YL\tw\No^*\right)
            +\sqrt{2}\lt\Rtt\Nf^*
          \right)\, ,\nonumber\\
        \Amat &=& \frac{1}{\sqrt{2}} \left(
            \sqrt{2}\lt\Rot\Nf
            -\YRt\tw\Rtt\No
            \right)\, ,\nonumber\\
        \Apib &=& \Rob\Uo^*-\lb\Rtb\Ut^*\, ,\nonumber\\
        \Amib &=& -\lt\Rob\Vt\, ,\nonumber\\
        \Apab &=& -\frac{1}{\sqrt{2}} \left(
          \Rob\left(\Nt^*-\YL\tw\No^*\right)
          -\sqrt{2}\lb\Rtb\Nth^*
        \right)\, ,\nonumber\\
        \Amab &=& -\frac{1}{\sqrt{2}} \left(
          -\sqrt{2}\lb\Rob\Nth
          +\YRb\tw\Rtb\No
          \right) \, ,
     \end{eqnarray}
with $Y_{L}$ and $Y_{R}^{t,b}$ the weak hypercharges of the left-handed $SU(2)_{L}$ doublet 
and right-handed singlet fermion, and $\lambda_{t}$ and $\lambda_{b}$
are the Yukawa couplings.

In the effective description approach, following hints from Higgs-boson physics
~\cite{Hall:1993gn, Carena:1994bv, Carena:1999py, Guasch:2001wv, Guasch:2003cv}, 
an effective Yukawa coupling is defined as:
\begin{equation}
\lambda_{b}^\text{eff}\equiv \frac{m_{b}^\text{eff}}{v_{1}}\equiv \frac{m_{b}(Q)}{v_{1}(1+\Delta m_{b})} \,, \hspace{0.5cm}
\lambda_{t}^\text{eff}\equiv \frac{m_{t}^\text{eff}}{v_{2}}\equiv \frac{m_{t}(Q)}{v_{2}(1+\Delta m_{t})} \,,
\label{eq:hbhbteff1}
\end{equation}
with $m_{q}(Q)$ $(q\equiv b, t)$ being the running quark mass and 
$\Delta m_{q}$ is the finite threshold correction. The SUSY-QCD contributions to $\Delta m_{q}$ are
\begin{eqnarray}
\label{eq:deltamq}
\Delta m_{b}^\text{SQCD}=\frac{2\alpha_{s}}{3\pi}\mg\mu\tan\beta\, I(m_{\sbottom_{1}},m_{\sbottom_{2}},\mg) \,, \nonumber \\
\Delta m_{t}^\text{SQCD}=\frac{2\alpha_{s}}{3\pi}\mg\frac{\mu}{\tan\beta}\, I(m_{\stopp_{1}},m_{\stopp_{2}},\mg) \,,
\end{eqnarray}
where $I(a,b,c)$ is the scalar three-point function at zero momentum transfer,
\begin{equation}
I(a,b,c)=\frac{a^{2}b^{2}\ln{(a^{2}/b^{2})}+b^{2}c^{2}\ln{(b^{2}/c^{2})+a^{2}c^{2}\ln{(c^{2}/a^{2})}}}
{(a^{2}-b^{2})(b^{2}-c^{2})(a^{2}-c^{2})} \,.
\end{equation} 
The effective description of the squark interaction consists in replacing the
tree-level quark masses in the couplings defined in Eqs.~(\ref{eq:V1Apm}) by the effective
Yukawa couplings of Eq.~(\ref{eq:hbhbteff1}), and use this Lagrangian to
compute the partial decay width (see~\cite{Guasch:2008fs} for details). 
A \textit{Yukawa-improved decay width computation} has been
defined in~\cite{Guasch:2008fs} and it showed that the effective description 
using just the Yukawa threshold corrections of Eq.~(\ref{eq:hbhbteff1}) 
is not enough for the squark partial decay widths description. 
The one-loop corrections develop a term which grows
as the gluino mass $\mg$~\cite{Hikasa:1995bw}, 
which is absent in the effective Yukawa couplings in Eq.~(\ref{eq:hbhbteff1}).
Therefore, the QCD corrections to squark partial decay widths produce explicit
non-decoupling terms of the sort $\log \mg$.
To understand those terms a renormalization group analysis is in 
order~\cite{Guasch:2008fs}. It is possible to construct an
effective theory below the gluino mass scale, which contains only
squarks, quarks, charginos, neutralinos and gluons in the light sector
of the theory, and integrate out the gluino contributions. We calculate
the renormalization group equations (RGE) of the gaugino and higgsino
couplings, and perform the matching with the full MSSM couplings at the
gluino mass scale $\mg$. Only the logarithmic RGE effects have been
considered, neglecting the possible threshold effects at the gluino mass scale.
Since the effective theory does not contain
gluinos, only the contributions from the gluon have to be taken into
account. In~\cite{Guasch:2008fs}, they showed that the effective
description~(\ref{eq:logterms}) approximates the full one-loop
computation to within $2-5\%$ for large enough gluino masses
($\mg\gtrsim 1\TeV$). The effects of the new 
logarithmic terms are more visible in the gaugino-like channels, where the
Yukawa couplings play no role, and the bulk of the corrections
corresponds to the log terms. In the
higgsino-like channels their importance is less apparent.

Finally, a simple expression for 
the effective description of squark/chargino/neu\-tra\-lino couplings is given
by~\cite{Guasch:2008fs,Abrahantes:2012xm}:
\begin{eqnarray}
g^\text{eff}(Q)&=&g\,\left(\frac{\al(Q)}{\al(\mg)}\right)^\frac{2}{\beta_{0}}\simeq g\left( 1-\frac{\al(Q)}{\pi}\log  \frac{Q}{\mg}\right) \,, \nonumber\\
\tilde{\lambda}_{b,t}^\text{eff}(Q)&=&\lambda_{b,t}^\text{eff}(Q)\,\left(\frac{\al(Q)}{\al(\mg)}\right)^\frac{-2}{\beta_{0}}
\simeq\lambda_{b,t}^\text{eff}(Q)\left( 1+\frac{\al(Q)}{\pi}\log \frac{Q}{\mg}\right) \,,
\label{eq:logterms}
\end{eqnarray} 
where $\beta_0$ is the QCD $\beta$-function and 
$\lambda^\text{eff}(Q)$ are the effective Yukawa couplings,
Eq.~(\ref{eq:hbhbteff1}).
Then, the effective description of squark interactions consist of replacing the 
tree-level quark masses and/or gaugino and higgsino couplings in
Eq.~(\ref{eq:V1Apm})
by the effective couplings as giving above, Eq.~(\ref{eq:logterms}).

After introducing these expressions in computer codes as {\tt MadGraph}, 
a good description for squark
decays into charginos and neutralinos is accomplished, and then
we are able to
compute any physical processes involving these vertices,
including the leading radiative corrections.

\section{New physics $t\bar{t}$ observables}
\label{observables}

It is known that by studying the final states
with top quarks at the LHC and measuring the top-quark polarization we are able to
differentiate the allowed MSSM scenarios. One possible scenario to explore 
is the case of the production of two top-squarks decaying into neutralinos and tops, 
$pp\to \tilde t_i \tilde t_i^\ast \to t\bar{t} \chi \chi$, where $\chi$ stands for $\tilde \chi_1^0$, $\tilde \chi_2^0$
and $\tilde t_i$ for $\tilde t_1$, $\tilde t_2$.

The resulting polarization of the top quark coming from the top-squark decays, 
$\mathcal{P}_{t}$, reads as~\cite{Belanger:2013gha}:
\begin{equation}
\mathcal{P}_{t}=\frac{[(A_{-\alpha i}^{(t)})^{2}-(A_{+\alpha i}^{(t)})^{2}]f_{1}}{(A_{-\alpha i}^{(t)})^{2}+(A_{+\alpha i}^{(t)})^{2}-2A_{-\alpha i}^{(t)}A_{+\alpha i}^{(t)}f_{2}},
\label{eq:polarization}
\end{equation}
where $f_{1,2}$ are pure kinematic factors given by
\begin{equation}
f_{1}=m_{t}\,\frac{p_{\chi}\cdot s_{t}}{p_{t}\cdot p_{\chi}},\ f_{2}=m_{t}\,\frac{m_{\chi}}{p_{t}\cdot p_{\chi}}\,,
\label{eq:polarizationFactors}
\end{equation}
with $m_{t}$, $p_{t}$ and $s_{t}$ denoting the top mass, momentum and longitudinal spin vector, respectively, and $p_{\chi}$
and $m_{\chi}$ the neutralino momentum and mass, and $A_{\pm\alpha i}^{(t)}$ are the neutralino
couplings defined in~Eq.(\ref{eq:V1Apm}). In the rest frame of the decaying particle these factor reduce to
\begin{equation}
f_{1}=\frac{\lambda^{\frac{1}{2}}(m_{\tilde t}^{2}, m_{t}^{2},m_{\chi}^{2})}{m_{\tilde t}^{2} - m_{t}^{2}-m_{\chi}^{2}},\ 
f_{2}=\frac{2m_{t}m_{\chi}}{m_{\tilde t}^{2} - m_{t}^{2}-m_{\chi}^{2}}
\label{eq:polarizationFactorsRest}
\end{equation}
and $\lambda(x,y,z)=x^{2}+y^{2}+z^{2}-2xy-2yz-2xz$.
By means of these expressions the polarization 
of the top quarks can be calculated at tree level and also in the 
effective approximation of squark interactions.

The measured particle assessing top-quark polarization is the electron
coming from the semi-leptonic decay of the $W$ boson from 
$t\to bW \to b l \nu_{l}$~\cite{Boos:2003vf,Godbole:2010kr,Godbole:2011vw,Prasath:2014mfa}. 
The top-quark polarization enters the lepton angular distribution in the following way~\cite{Godbole:2010kr}
  \begin{eqnarray}
 \frac{1}{\Gamma_{l}}\frac{d\Gamma_{l}}{d\cos\theta_{l}}=\frac{1}{2}(1-\beta_{t}^{2}) \, (1-\mathcal{P}_{t}\beta_{t}) \,
 \frac{1+\frac{\mathcal{P}_{t}-\beta_{t}}{1-\mathcal{P}_{t}\beta_{t}}\cos\theta_{l}}{(1-\beta_{t}\cos\theta_{l})^{3}}\,,
 \label{eq:differentialDistLept}
\end{eqnarray}
where $\theta_{l}$ is the angle between the top-quark and the lepton directions in the laboratory frame, 
and $\beta_{t}$ is the top-quark velocity:
\begin{equation}
\beta_{t}=\frac{|p_{t}|}{E_{t}}\,,
\label{eq:topboost}
\end{equation}
being $E_{t}$ the total energy of the top quark. The effects from original top-quark polarization, $\mathcal{P}_{t}$, are entangled with the boost in the form
 of an effective polarization~\cite{Godbole:2010kr}
 \begin{equation}
 \mathcal{P}_{t}^\text{eff}=
\frac{\mathcal{P}_{t}-\beta_{t}}{1-\mathcal{P}_{t}\beta_{t}}\,.
 \label{eq:effPol}
 \end{equation}
To measure the top-quark polarization one can define an asymmetry in $\theta_{l}$.
Because the $\theta_{l}$ distribution is non-symmetric, one has the freedom to define 
asymmetries with respect to different angles, for example~\cite{Belanger:2012tm}
\begin{equation}
A_{\theta_{l}}=\frac{\sigma(\theta_{l}<\pi/4)-\sigma(\theta_{l}>\pi/4)}{\sigma(\theta_{l}<\pi/4)+\sigma(\theta_{l}>\pi/4)} \, \, ,
\label{eq:asymobserv2}
\end{equation}
where $\sigma$ is the integrated cross-section.

Another option to characterize the asymmetry is using the azimuthal angle.
We define the following axes system: the $\hat{z}$-axis is defined by the proton direction, and the 
$\hat{x}-\hat{z}$ plane is defined by the top-quark direction and the $\hat{z}$-axis, then
$\phi_{l}$ is the azimuthal angle of the lepton in this system.
Because at the LHC the initial state has identical particles, the $\hat{z}$-axis can point in the direction of either 
proton, and it is not possible to distinguish between $\phi_{l}$ and $2\pi-\phi_{l}$~\cite{Belanger:2013gha}. 
We can relate $\theta_{l}$ to $\phi_{l}$ by using the spherical angles coordinates in this axes system: the top-quark
angular variables are $(\hat{\theta}_{t},\,  \hat{\phi}_{t}=0)$ and the lepton ones are $(\hat{\theta}_{l},\,  \hat{\phi}_{l}=\phi_{l})$, then:
\begin{equation}
\cos\theta_{l}=
\cos\hat{\theta_{t}}\cos\hat{\theta}_{l}+\sin\hat{\theta}_{t}\sin\hat{\theta}_{l}\cos\phi_{l}\, ,
 \end{equation}
 in this way the lepton distribution~\eqref{eq:differentialDistLept}, after integrating over $\hat{\theta}_{t}$ and $\hat{\theta}_{l}$,
 picks up a $\phi_{l}$ dependence. We define then the asymmetry~\cite{Belanger:2012tm}
 \begin{equation}
A_{\phi_{l}}=\frac{\sigma(\cos\phi_{l}>0)-\sigma(\cos\phi_{l}<0)}{\sigma(\cos\phi_{l}>0)+\sigma(\cos\phi_{l}<0)} \, ,
\label{eq:asymobserv1}
\end{equation}
where $\sigma$ is the integrated cross-section.

Besides, $\beta_{t}$, $A_{\phi_{l}}$ and $A_{\theta_{l}}$, there are
other top-quark-wise observables characterizing the events, such as:
\begin{eqnarray}
z&=&\frac{E_{b}}{E_{t}}\,, \label{eq:zfrac} \\
u&=&\frac{E_{l}}{E_{l}+E_{b}}\,. \label{eq:ufrac}
\end{eqnarray}
The ratios $z$ and $u$, with $E_t$, $E_b$ and $E_l$ being the lab-frame
energies of the top-quark, and the bottom-quark and the lepton coming
from its decay, respectively, are sensitive to the top-quark polarization
when the top-quarks are highly boosted~\cite{Shelton:2008nq}.
The distributions of these variables can be explored using the {\tt MadGraph}
implementation containing the effective squark 
approximation~\cite{Abrahantes:2012xm}, allowing
us to acknowledge whether top-quark polarization observables are sensitive to 
the effective approximation of squark interactions. 
It worth mentioning that for our new physics scenario the top-quark polarization
is further affected by the underlying event kinematics: our top-quark came from an already boosted system, the squark.
Thus, the resulting top-quark polarization has a dependence on the squark boost~\cite{Belanger:2012tm}.  

Finally, it is known that the top-quark production at the Tevatron is dominated by the $q\bar{q}$ annihilation, 
hence the charge asymmetry will be reflected
not only in the partonic rest frame but also in the center of mass system of proton and antiproton. The situation is more complex
for proton-proton collisions at the LHC, where no preferred direction is at hand in the laboratory frame, thus 
 lacking a natural definition for the charge asymmetry given the symmetric 
nature of the incoming protons. However, the parton distributions inside the protons are not 
symmetric for quarks (mainly valence quarks) and antiquarks (all sea quarks), meaning quarks  
usually carry more momentum than antiquarks. 
For a positive (negative) charge asymmetry in $q\bar{q}\to t\bar{t}$ events, 
the top-quark (top-antiquark) is more likely to be produced in the 
direction of the incoming quark in the $t\bar{t}$ rest frame, resulting in a broader 
(narrower) rapidity distribution of top-quarks than of top-antiquarks in the laboratory frame. The difference in 
the absolute values of the rapidities ($y$) of the top-quarks and antiquarks, $\Delta|y_{t}|=|y_{t}|-|y_{\bar{t}}|$, is 
therefore a suitable observable to measure the $t\bar{t}$ charge asymmetry. 
Several processes beyond the SM can alter this asymmetry
~\cite{AguilarSaavedra:2011hz, AguilarSaavedra:2011vw, Antunano:2007da, Djouadi:2011aj,
Ferrario:2008wm,Jung:2009jz,Shu:2009xf,Falkowski:2011zr,Falkowski:2012cu,Gripaios:2013rda}, either with vector or 
axial vector couplings or via interference with the SM. Hence the measurement of the charge asymmetry 
provides a useful tool to test for the presence of new physics that would be hidden in the $t\bar{t}$ invariant mass 
($m_{t\bar{t}}$) spectrum.

The charge asymmetry in $t \bar t$ production at the LHC can be defined as follows:
\begin{eqnarray}\label{eq:ACtt}
A_{C}^{t\bar{t}(l^{+}l^{-})}(\xi)&=&\frac{N(\Delta|\xi|>0)-N(\Delta|\xi|<0)}{N(\Delta|\xi|>0)+N(\Delta|\xi|<0)} \,,\label{eq:AC}
\end{eqnarray}
where $\Delta|\xi|=|\xi_{t(l^{+})}|-|\xi_{\bar{t}(l^{-})}|$ and 
$\xi$ is $\eta$ or $y$, the pseudo-rapidity ($\eta=-\log \tan\frac{\theta}{2}$) or rapidity 
($y=\frac{1}{2}\ln\frac{E+p_{z}}{E+p_{z}}$) of top-quarks and its semi-leptonic decay 
products, respectively. 
Ref.~\cite{Kuhn:2011ri} argued that since most of the charge asymmetry is
 concentrated at large rapidities, the statistical significance of any 
measurement will be enhanced, if the sample is restricted
 to larger rapidities. Therefore, a complementary asymmetry is defined by:
\begin{equation}
A_{t\bar{t}}(Y_\text{cut})=A_{C}^{t\bar t}(y)=
\frac{N(|y_{t}|>|y_{\bar{t}}|)-N(|y_{\bar{t}}|>|y_{t}|)}{N(|y_{t}|>|y_{\bar{t}}|)+N(|y_{\bar{t}}|>|y_{t}|)} \,\mbox{  with  } 
\frac{|y_{t}+y_{\bar{t}}|}{2} > Y_\text{cut}\, \, .
\label{eq:ACymean}
\end{equation}

Moreover, a kinematic cut on
\begin{equation}
\beta_{z,t\bar{t}}=\frac{|p_{z}^{t} + p_{z}^{\bar{t}}|}{E^{t}+E^{\bar{t}}}
\end{equation}
can be used to
enlarge the fraction of $q\bar q$ events, $\sigma(q\bar{q})$,
contained in the total cross-section, $\sigma_{total}=\sigma(q\bar{q})+\sigma(gg)$ 
(see, for example~\cite{AguilarSaavedra:2011cp}).

These asymmetries have been measured at the $\sqrt{s}=7\TeV$ LHC. The SM prediction for them is
around $\sim10^{-2}$~\cite{Kuhn:2011ri}. Current LHC experiments are not sensitive enough to measure a non-null asymmetry,
their results being compatible with the SM and zero at one standard deviation:
\begin{eqnarray}
A_{C}^{t\bar t} &=& 0.006\pm 0.010\, \, \mbox{\cite{Aad:2013cea} ,}\nonumber\\
&=&-0.010\pm0.019\, \, \mbox{\cite{Chatrchyan:2014yta} ,}\nonumber\\
&=& 0.018\pm 0.022\, \, (m_{t\bar{t}}>600\GeV) \mbox{~\cite{Aad:2013cea} ,}\nonumber\\
&=&0.011\pm0.018\, \, (\beta_{z,t\bar{t}}>0.6) \mbox{~\cite{Chatrchyan:2014yta} ,}\nonumber\\
A_{t\bar t}(Y_{\text{cut}}=0.7)&=& 0.015\pm 0.025\, \, \mbox{\cite{Aad:2013cea}}\, \, .\nonumber
\end{eqnarray}
The leptonic asymmetries are also compatible with zero but at two standard deviations:
\begin{eqnarray}
A_{C}^{l^{+}l^{-}} &=& 0.023\pm 0.014\, \, \mbox{\cite{ATLAS:2012sla} ,}\nonumber\\
&=&0.009\pm0.012\, \, \mbox{\cite{Chatrchyan:2014yta}}\, \, .\nonumber
\end{eqnarray}
The new physics models trying to explain the deviation found at Tevatron must have the complementary check against LHC 
measured asymmetries.

\section{Numerical analysis}
\label{num-results}

\begin{table}[t]
\begin{center}
\begin{tabular}{|c|c|c|c|}
\hline
Parameter&Description&Present\\
\hline
$G_{F}$&Fermi constant&$1.1663787\times 10^{-5}$ GeV$^{-2}$\\
$1/\alpha_{em}(m_{Z})$ &Inverse of Electromagnetic coupling constant&137.035999074\\
$\alpha_{s}(m_{Z})$ &Strong coupling constant&0.1184\\
$m_{top} $&top-quark mass &173.5 GeV\\
$m_{b}^{\overline{MS}}$ &bottom-quark mass&4.18 GeV\\
$m_{Z}$&Z boson mass&91.1876 GeV\\
$m_{W}$&W boson mass&80.385 GeV\\
\hline
\end{tabular}
\caption{Parameters of the Standard Model as in~\cite{Beringer:1900zz}.} 
\label{tab:SMparam}
\end{center}
\end{table}
We present the numerical analysis for fixed values of the
SUSY parameters and make plots by changing one parameter at a time. 
However, we stress that our programs are able to perform 
computations for any MSSM
parameter space point and they admit SLHA~\cite{Skands:2003cj} input for
easy interaction with other programs/routines. For the SM parameters\footnote{The most recent value for the 
top-quark mass is $173 \pm 0.3 \pm 0.7$ GeV (value $\pm$ stat $\pm$ syst)~\cite{TOP2014}.
This value is very close to the ones used in this work and it has no any consequence in our results.} we
use those given in~\cite{Beringer:1900zz}, listed in
table~\ref{tab:SMparam}. 
For the SUSY parameters, we choose four
different scenarios summarized in table~\ref{tab:allMSSM}. 
Here $M_1$ and $M_2$ are the $U(1)$ and $SU(2)$ gaugino mass parameters; respectively, 
$m_{\tilde{g}}$ is the gluino mass, 
$A_f (f=t,b,\tau)$ denotes the trilinear Higgs-quark coupling, 
$\mu$ is the higgsino mass parameter,
$M_{A}$ is the pseudoscalar mass, $\tan \beta$ 
is the ratio between the Higgs fields vacuum expectation values 
and the last six rows correspond with the soft-SUSY breaking parameters 
in the squark sector. On one side, because of the
comparison with previous results and for illustrative purposes,
we choose a parameter set defined as {\it Def}~\cite{Abrahantes:2012xm}.
This scenario has been largely explored in the above reference within a
very good accuracy of the effective approximation, namely that 
the effective approximation provides 
a good description of the radiative-corrected squark partial decay 
widths if the gluino mass is heavier than the top-squark mass. 
On the other side, SUSY parameters are
chosen from a modification of the $m^{h}_{max}$ scenario 
as in~\cite{Carena:2013qia}  
with negative trilinear couplings ($\modmin$), and 
the model 100267 from~\cite{Cahill-Rowley:2013gca} (\pMSSM).
Finally, we define a new scenario with relatively small masses for the 
squarks of the third generation in order to provide a scenario with top-squarks
capable of being produced in the next run of the LHC and compatible with
the effective approximation used along this work. We denote this 
parameter choice {\it LS} in our numerical analysis, corresponding 
with a scenario with {\it{light squark}}. 
Even if we present the input parameters for all the scalar sector, 
we restrict ourselves in the numerical analysis to the case of third generation 
squarks. With these input parameters, the central values for the physical 
SUSY particle masses are given in table~\ref{tab:massesMSSM}. 

\begin{table}[t]
\begin{center}
\begin{tabular}{|c|cccc|}
\hline
Parameter &{\it Def}&\modmin&\pMSSM &{\it LS}\\
\hline
$M_{1}$&  95.6&95.6&1018&310.9\\
 $M_{2}$&  200&200&2462&650\\
$\mg$&  3000&1500&3368&3000\\
 $A_{t}$& 1630&-1890&3793&1700\\
 $A_{b}$& 1630&-1890&-1285&1700\\
$A_{\tau}$& 1630&-1890&3827&0\\
$\mu$&  300&200&1911&800\\
$\tb$&   10&20&43.02&10\\
 $M_{A}$&500&700&3002&495.5\\
 $M_{\tilde{L}_{1,2}}$& 1000&500&1115&1000\\
 $M_{\tilde{L}_{3}}$& 1000&1000&1086&1000\\
 $M_{\tilde{E}_{1,2}}$& 1000&500&2554&1000\\
 $M_{\tilde{E}_{3}}$& 1000&1000&2408&1000\\
 $M_{\tilde{q}_{1,2}}$& 800&1500&1100&1000\\
 $M_{\tilde{q}_{3}}$& 800&1000&1624&700\\
 $M_{\tilde{U}_{1,2}}$& 1000&1500&2604&1000\\
 $M_{\tilde{U}_{3}}$&  1000&1000&2829&1000\\
  $M_{\tilde{D}_{1,2}}$& 1000&1500&3156&1000\\
 $M_{\tilde{D}_{3}}$& 1000&1000&3024&1000\\
\hline
\end{tabular}
\caption{SUSY parameters for the MSSM scenarios: {\it Def}~\cite{Abrahantes:2012xm},
$\modmin$~\cite{Carena:2013qia}, model 100267~\cite{Cahill-Rowley:2013gca} (\pMSSM) and {\it LS}.
Masses and trilinear couplings are in \GeV.}
\label{tab:allMSSM}
\end{center}
\end{table}

\begin{table}[t]
\begin{center}
\begin{tabular}{|c|cccc|}
\hline
Masses&{\it Def}&\modmin&\pMSSM&{\it LS}\\
\hline
$\mg$&3000&1500&3368&3000\\
\hline
$m_{\stopp_{1}}$&718.8&835.7&1609&626.9\\
$m_{\stopp_{2}}$&1086&1165&2848&1074\\
\hline
$m_{\tilde \chi^0}$&91.7, 176.3 - 334.6&87.9, 151.4 - 266.4&1017, 1906-2469&309.3, 629.5 - 824.6\\
$m_{\tilde \chi^{+}}$&175.3, 334.9&147.5 - 266.8&1905, 2469&629.3, 824.2\\
\hline
$m_{h}$&125.3&127.1&127.8&125.6\\
\hline
\end{tabular}
\caption{Masses of SUSY particles, in $\GeV$.}
\label{tab:massesMSSM}
\end{center}
\end{table}
At present the ATLAS and CMS collaborations have already put some stringent
limits~\cite{Chatrchyan:2013wxa,Chatrchyan:2013xna,Aad:2013ija,Chatrchyan:2013mya,Aad:2014qaa,Aad:2014mha,Aad:2014pda,Khachatryan:2014doa}
on the masses of gluino and squarks, specially of the first and second generations.
Our top-squark mass parameter choices are
larger than the excluded ones, moreover the exclusion limits would be
loosened by allowing the existence of several decay channels for the
top squark. Besides, since we are interested in the effective description of squark
interactions, the gluino mass is chosen preferably large to enhance the effects of the 
logarithmic terms. Note also that if the gluino decay channel is open,
it will be the dominant decay channel for squarks, rendering the
chargino/neutralino channels phenomenologically irrelevant. Therefore
our region of interest is:
\begin{equation}
\mg+m_q > \msq \,\,.
\label{eq:gluinocondition}
\end{equation}
The analysis of the accuracy of the effective approximation was
performed in~\cite{Guasch:2008fs}, using the parameter set
\textit{Def} as an example. We have
additionally checked that the same conclusions hold for the MSSM 
scenarios analyzed in the present work, namely that the effective
approximation provides a good description of the radiative-corrected
partial decay widths, if the gluino mass is heavier than the
top-squark mass ($\mg\gtrsim 1000\GeV$).
We have also checked that our SUSY parameters sets are compatible with
present values of the Higgs boson mass~\cite{Aad:2012tfa,Chatrchyan:2012ufa}.
We use a self-coded routine containing the expressions 
of~\cite{Heinemeyer:1999be} for
the computation of the Higgs boson mass at two-loop level. 
The results for the mass of the 
lightest CP-even Higgs boson mass for each scenario are also included 
in table~\ref{tab:massesMSSM}. 

\begin{figure}[t]
\centering
\begin{tabular}{cc}
\includegraphics[height=3cm,angle=0]{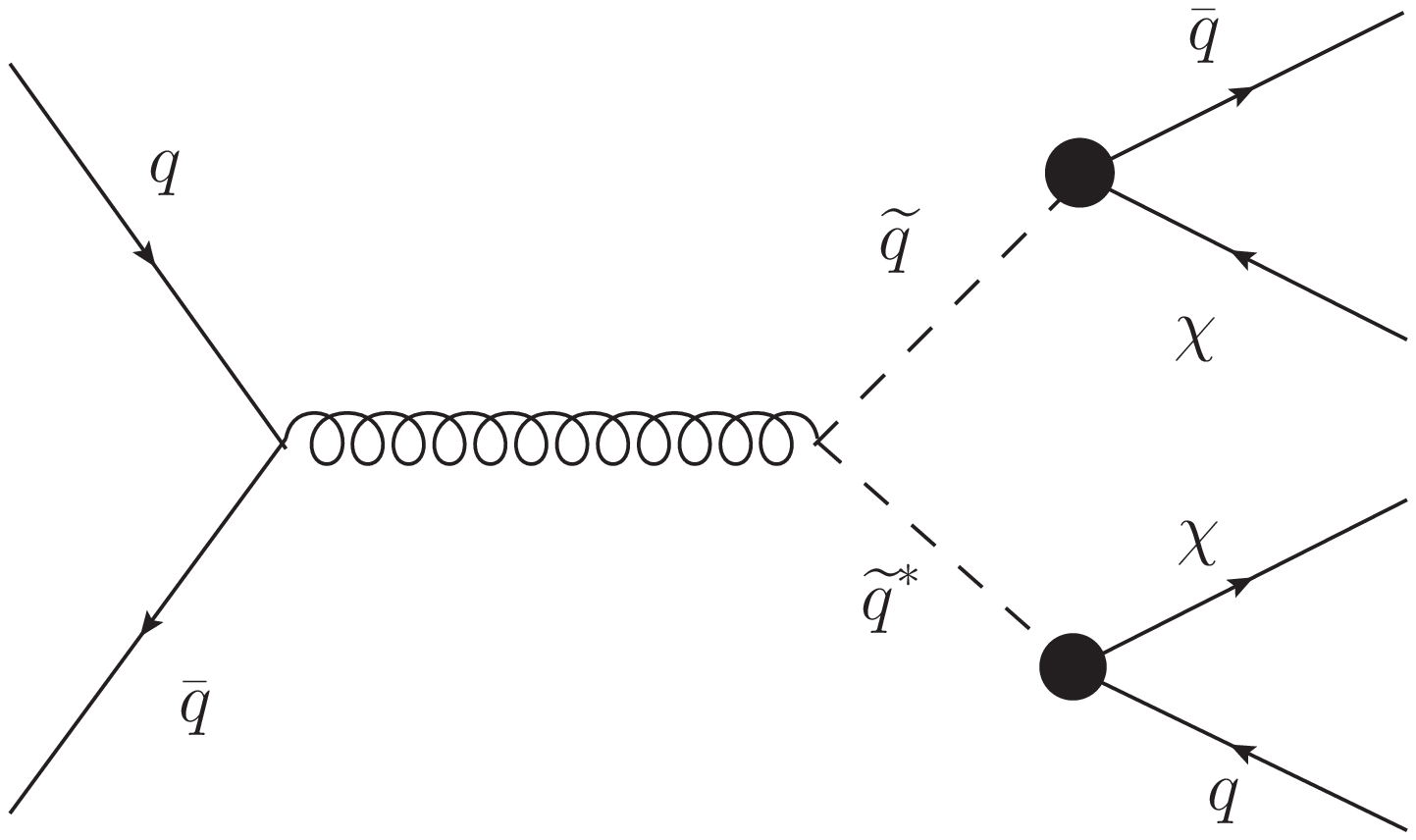}&
\includegraphics[height=3cm,angle=0]{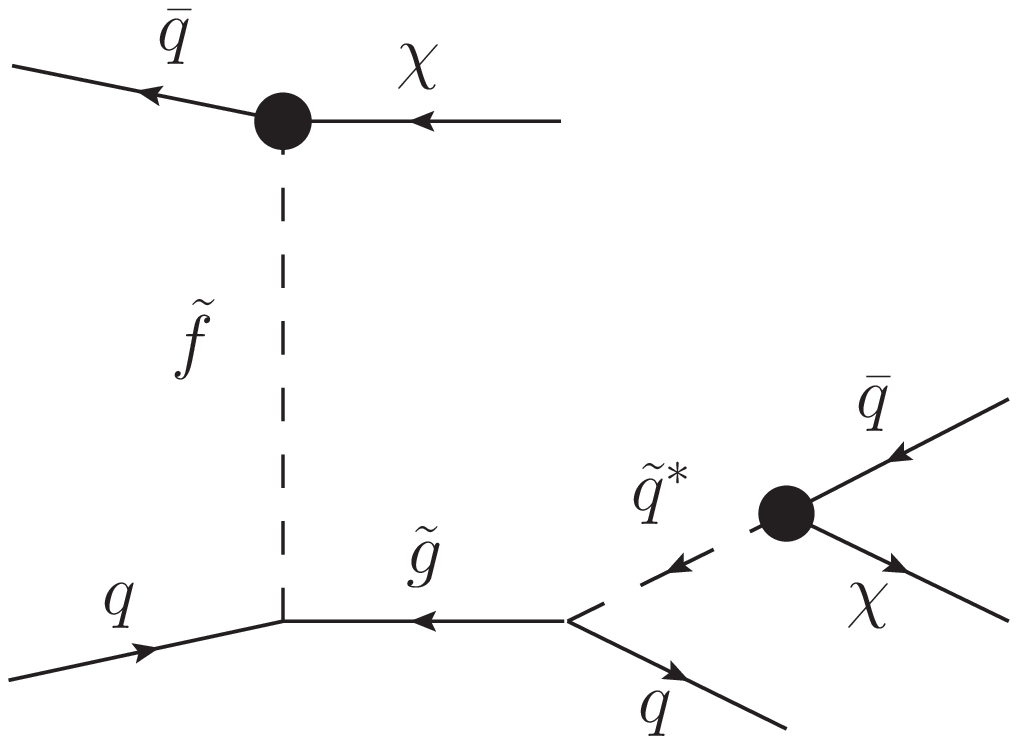}\\
(a)&(b)\\
\end{tabular}
\caption{Generic Feynman diagrams contributing to
$\sigma(q\bar{q}\rightarrow ({q}\chi)({\bar{q}}\chi))$\, \textbf{(a)}
double resonant diagrams\, \textbf{(b)} single resonant diagrams.}
\label{fig:pairdiag}
\end{figure}

Our aim is the phenomenological analysis of the MSSM contributions to the top-quark
charge asymmetries and top-quark polarization at the LHC. 
In particular, we concentrate in the contributions coming from the decays of the top-squarks
($\tilde t_{1}$ and $\tilde t_{2}$) into top quarks 
plus missing energy, both at tree level and at one-loop order.
Figure~\ref{fig:pairdiag} shows the generic Feynman diagrams contributing the most to 
$\sigma(q\bar{q}\rightarrow
({q}\chi)({\bar{q}}\chi))$. Here $\tilde{f}$ denotes the squarks of 
the first and the second generations. 
Left panel of figure~\ref{fig:pairdiag} includes the generic 
double resonant diagrams ($\sigma(q\bar{q}\to\squark_a\squark_a^*\rightarrow
({q}\chi)({\bar{q}}\chi))$) and the right panel 
shows the single resonant diagrams.
Regarding the simulation procedure, we have generated 5 millions of events,
by means of {\tt MadGraph}, for each one of the four SUSY scenarios described
in table~\ref{tab:allMSSM} and also 5 millions of events for the $t \bar t$ SM background,
computed at NLO.

First of all, we have computed the top-squark partial decay widths in
each scenario, focusing on reactions where both $\tilde t_1$ and 
$\tilde t_2$ decay
into a top-quark and the lightest neutralino $\tilde \chi_1^0$. Results 
for all the branching ratios 
at tree level are presented in table~\ref{tab:BRMSSM}. Note that 
the more convenient MSSM scenarios for 
our purpose are  $\pMSSM$ and {\it LS}. 
In these two cases BR$(\stopp_{1}\to t \tilde \chi_1^0)$ is maximal, 
the other $\tilde t_1$ decays channels being closed. 
For $\tilde t_2$ all decays channels are open being 
BR$(\stopp_{2}\to X)$ ($X\equiv \stopp_{1} h, \stopp_{1} Z^0, \tilde b_{1} W^+$)
maximal. In the  {\it Def} and $\modmin$ scenarios there exist other 
allowed decay channels for the top-squarks and the 
$\stopp_{1,2}\to t \tilde \chi_1^0$ are suppressed.
However, we choose these two other MSSM sets to have a more 
complete phenomenological analysis and
to arrive to general conclusions. 

\begin{table}[t]
\begin{center}
\begin{tabular}{|l|c|c|c|c|c|}
\hline
Branching ratio&{\it Def}&\modmin&\pMSSM&{\it LS}\\ \hline
BR$(\stopp_{1}\to t \tilde \chi_1^0)$&0.011&0.057&1&1\\
BR$(\stopp_{1}\to t \tilde \chi_2^0)$&0.073&0.148&0&0\\
BR$(\stopp_{1}\to t \tilde \chi_3^0)$&0.186&0.296&0&0\\
BR$(\stopp_{1}\to t \tilde \chi_4^0)$&0.352&0.058&0&0\\
BR$(\stopp_{1}\to b \tilde \chi_1^+)$&0.125&0.385&0&0\\
BR$(\stopp_{1}\to b \tilde \chi_2^+)$&0.251&0.056&0&0\\ \hline
BR$(\stopp_{2}\to t \tilde \chi_1^0)$&0.030&0.010&0.029&0.030\\
BR$(\stopp_{2}\to t \tilde \chi_2^0)$&0.048&0.045&0.051&0.019\\
BR$(\stopp_{2}\to t \tilde \chi_3^0)$&0.151&0.146&0.054&0.041\\
BR$(\stopp_{2}\to t \tilde \chi_4^0)$&0.071&0.194&$\sim10^{-5}$&0.008\\
BR$(\stopp_{2}\to b \tilde \chi_1^+)$&0.107&0.028&0.109&0.048\\
BR$(\stopp_{2}\to b \tilde \chi_2^+)$&0.139&0.258&$\sim10^{-5}$&0.042\\
BR$(\stopp_{2}\to X)$&0.454&0.319&0.757&0.812\\
\hline
\end{tabular}
\caption{Branching ratios of $\stopp_{1,2}$ decays for
  different MSSM scenarios, at tree level.
 $\stopp_{2}\to X$ stands for the sum of branching of all other possible $\stopp_{2}$ decay channels.}
\label{tab:BRMSSM}
\end{center}
\end{table}

The SUSY-QCD contributions we are interested in are
\begin{equation}
pp\to t \bar{t} \tilde \chi_1^0 \tilde \chi_1^0 \to  
b \,\bar{b}\, l\,\nu_{l}\,l^{\prime}\,\nu_{l^{\prime}}\, \tilde \chi_1^0 \tilde \chi_1^0 \,,
\end{equation}
where $p$ implies $g,u,d,s,c$ initial states 
and $l$ resume $e$ and $\mu$ leptons. 
Besides, leptons in the final state appear only as product 
of the top-quark decay chain $t\to bW \to b l \nu_{l}$.
We apply kinematic cuts on the transverse momentum $p_{T}$
and pseudo-rapidity $\eta$ of the final leptons and $b$ quarks,
being $p_{T}>20$ GeV and  $|\eta|<2.5$.

Relative deviation, $\delta$ in \%, of any quantity, R, is calculated as:
\begin{equation}
\delta=\frac{R^\text{eff}-R^\text{tree}}{R^\text{tree}}\times 100\,.
\end{equation}

\begin{table}[t]
\begin{center}
\begin{tabular}{|r|cc|cc|cc|}
\hline
&\multicolumn{2}{c|}{$\sigma(pp\to b\bar{b}ll^{\prime}+E_T^\text{miss})[\fb]$}&
\multicolumn{2}{c|}{$\kappa_{p_T^\text{miss} > 200 \, \text{GeV}}$}&
\multicolumn{2}{c|}{$\kappa_{p_T^\text{miss} > 300 \, \text{GeV}}$}\\
\hline
SM &\multicolumn{2}{c|}{$18.1\times10^{3}$}&\multicolumn{2}{c|}{$1.02 \times 10^{-2}$}&
\multicolumn{2}{c|}{$1.62 \times 10^{-3}$}\\
\hline
MSSM &Tree&Eff&$\>\>\>\>$Tree$\>\>\>\>$&Eff&$\>\>\>\>$Tree$\>\>\>\>$&Eff\\
\hline
${\it Def}$&$2.39\times10^{-4}$&$4.52\times10^{-4}$&0.76&0.75&0.55&0.53\\
\modmin&$2.32\times10^{-3}$&$3.26\times10^{-3}$&0.79&0.79&0.59&0.59\\
\pMSSM&$2.76\times10^{-3}$&$2.76\times10^{-3}$&0.81&0.81&0.62&0.62\\
{\it LS}&$2.72$&$2.77$&0.40&0.40&0.12&0.12\\
\hline
\end{tabular}
\caption{Cross section of $pp\to b\,\bar{b}\,l\,l^{\prime}+E_T^\text{miss}$ 
in the SM ($E_T^\text{miss}=\nu_{l}\nu_{l^{\prime}}$) and in the 
MSSM ($E_T^\text{miss}=\nu_{l}\nu_{l^{\prime}}\tilde \chi_1^0\tilde \chi_1^0$) 
for studied SUSY scenarios at the $\sqrt{s}=14\TeV$ LHC. 
$\kappa$ is the fraction of events surviving a giving cut on $p_T^\text{miss}$.}
\label{tab:xsectionnew}
\end{center}
\end{table}

Table \ref{tab:xsectionnew} shows the results for the cross section 
$\sigma(pp\to b\bar{b}ll^{\prime}+E_T^\text{miss})$ in $\pb$
both in the SM ($E_T^\text{miss}=\nu_{l}\nu_{l^{\prime}}$) and in
the MSSM ($E_T^\text{miss}=\nu_{l}\nu_{l^{\prime}}\tilde \chi_1^0\tilde
\chi_1^0$) at the $\sqrt{s}=14\TeV$ LHC. 
We use {\tt MadGraph} for the computation.
The values of the cross section are given
in each scenario both at tree level and 
in the effective approximation. Besides, the results for the fraction
of surviving events above
certain cut on $p_T^\text{miss}$ defined as 
$p_T^\text{miss}=\sum_{i=\nu_{l},\tilde \chi_1^0} p_{T}^{i}$ are presented 
in the last two columns.
$\kappa$ is the fraction of events surviving a giving cut on $p_T^\text{miss}$. 
As expected, the cross-section in all the MSSM scenarios is suppressed with respect to the
SM one. The values of the MSSM cross-section are between 
$2 \fb$ and $10^{-4} \fb$, depending on the SUSY parameter choice. 
Clearly, the most promising scenario for this analysis is
{\it LS}, having a maximum value for the cross-section of about $2 \fb$. 
The radiative corrections can be large in some
scenarios. For example, the relative deviation $\delta$ between the tree-level
calculation and the effective results is around
$89\%$ and $40\%$ in the {\it Def} and \modmin scenarios, respectively. 
However, the situation change drastically for
the other two scenarios, being the above relative deviation equal or less than $1\%$.

\begin{table}[t]
\begin{center}
\begin{tabular}{|c|cc|}
\hline
\multicolumn{3}{|c|}{$LS$ signal significances}\\
\hline
Significances & $\mathcal{L}=300\fb^{-1}$ & $\mathcal{L}=1000\fb^{-1}$ \\
\hline
$S/B$ & 0.014 & 0.014 \\
$S/\sqrt{B}$ & 1.39 & 2.54 \\
$S/\sqrt{S+B}$ & 1.38 & 2.53 \\
$S/\sqrt{B+(0.2B)^2}$ & 0.07 & 0.07 \\
\hline
\end{tabular}
\caption{Signal significances for the $LS$ scenario, with $S = \sigma_S \mathcal{L}$ and $B = \sigma_B \mathcal{L}$.}
\label{tab:LSsignificances}
\end{center}
\end{table}
We complete the analysis presenting the results
for $\sqrt{s}=14\TeV$ and $\mathcal{L}=300\fb^{-1}$,
the anticipated integrated luminosity that will be delivered by
LHC in its first 10 years of life~\cite{Barletta:2013ooa},
or $\mathcal{L}=1000\fb^{-1}$ for the high-luminosity LHC (HL-LHC)~\cite{Jakobs:2011zz}.
The {\it LS} scenario has the best signal to background relations 
of all the scenarios presented in this work and its expected signal significances for
the highest cut on $p_T^\text{miss}$ are shown in table~\ref{tab:LSsignificances}.
There are some recent works very useful
to identify $K$-factors that could be applied to the MSSM cross sections:
a $K$-factor of 1.25-1.3~\cite{Germer:2014jpa} can be estimated at NLO for top-squark production
at the LHC~\cite{Germer:2014jpa,Hollik:2008yi,Hollik:2008vm,Germer:2010vn,Germer:2011an} and 
a $K$-factor of 1.8~\cite{Beenakker:2014sma} at next-to-next-to-leading-logarithmic (NNLL) 
accuracy~\cite{Beenakker:2014sma,Beenakker:2009ha,Beenakker:2010nq,Beenakker:2011fu,Beenakker:2011dk,Beenakker:2011sf,Beenakker:2013mva}
over the NLO cross section,
resulting in an overall $K$-factor of 1.4, which has been considered
for the calculations in table~\ref{tab:LSsignificances}.
We have also to take into account the $K$-factors for the SM which can be extracted from the
well-known NNLO+NNLL QCD corrections to $t \bar t$ production~\cite{Cacciari:2011hy,Baernreuther:2012ws,Czakon:2012zr,Czakon:2012pz,Czakon:2013goa},
estimating an approximate $K$-factor of 1.1.
The results for the HL-LHC in table~\ref{tab:LSsignificances} are close to the lower limits of signal 
observation (around 3) and whilst
they could be improved somehow if one applies some more cuts optimized to reduce the SM contribution,
like a cut on the top-quark polarizations,
it seems difficult to increase sufficiently the signal significances
to achieve values near 5, considered as the signal discovery.

In that sense, in order to try to find scenarios with more promising signal significances,
one can imagine simplified models~\cite{{ArkaniHamed:2007fw,Alwall:2008va,Alwall:2008ag,Alves:2010za,Alves:2011sq,Alves:2011wf}}
with only one neutralino and one stop at low energies,
whose masses should be much lighter than in the four scenarios considered in this work.
One example of this kind of simplified models could be a modified $LS$ scenario with $M_1 =$ 100 GeV
and $M_{\tilde q_3} =$ 375 GeV. This choice of soft-SUSY breaking parameters would give rise to a supersymmetric spectrum
with $m_{\tilde \chi_1^0} =$ 99.3 GeV and $m_{\tilde t_1} =$ 310 GeV. In this case, BR$(\tilde t_1 \to t \tilde \chi_1^0) =$ 1
and then we would obtain $\sigma(pp \to \tilde t_1 \tilde t_1 \to b \bar b l l^\prime + E_T^\text{miss}) \sim$ 150 fb at $\sqrt{s} = $ 14 TeV.
If we consider a similar fraction of surviving events around 10\% after applying cuts on $p_T^\text{miss}$ as in table~\ref{tab:xsectionnew}
for the $LS$ scenario, the resulting cross section would be $\sim 15$ fb.
Taking into account the $K$-factors mentioned above for the SM background and the MSSM scenarios,
the expected signal significances for this simplified model would be $S/B \sim$ 0.7 (clearly larger than ${\cal O}(10\%)$ which is
considered as the minimum value to have an observable signal) and $S/\sqrt{B+(0.2B)^2} \sim$ 3.4 for both total integrated luminosities
of 300 fb$^{-1}$ and 1000 fb$^{-1}$. These significances would really improve the results of table~\ref{tab:LSsignificances} and show
what class of scenarios could be testable at the LHC performing these search strategies, which are out of the scope of this work.
Anyway, the effects of the effective squark approximation presented and used here 
are practically negligible for this kind of simplified models, depending on the SUSY parameter space.

\subsection{Top-quark charge asymmetries}
\label{top-asymmetries}

\begin{table}[t]
\begin{center}
\resizebox{\textwidth}{!}{
\begin{tabular}{|c|cccccc|}
\hline
&\multicolumn{6}{c|}{$\beta_{z,t\bar{t}}>0$($\beta_{z,t\bar{t}}>0.6$)}\\
&\multicolumn{2}{c}{$100\times A_{C}^{t\bar{t}}(y)$}&\multicolumn{2}{c}{$100\times A_{C}^{t\bar{t}}(\eta)$}&\multicolumn{2}{c|}{$100\times A_{C}^{l^{+}l^{-}}(y)$}\\
\hline
SM $pp\to t\bar{t}$&\multicolumn{2}{c}{0.229(0.275)}&\multicolumn{2}{c}{0.307(0.350)}&\multicolumn{2}{c|}{0.195(0.214)}\\
\hline
$pp\to t\bar{t}\tilde \chi_1^0\tilde \chi_1^0$&Tree&Eff&Tree&Eff&Tree&Eff\\
\hline
{\it Def}&0.354(0.413)&0.191(0.152)&0.508(0.582)&0.332(0.269)&0.334(0.416)&0.202(0.141)\\
\modmin&0.196(0.165)&0.167(0.237)&0.364(0.384)&0.293(0.437)&-0.371(-0.881)&-0.316(-0.637)\\
\pMSSM&0.052(-0.069)&-0.003(-0.017)&0.059(-0.057)&-0.001(0.018)&-0.008(-0.114)&-0.026(-0.122)\\
LS&0.013(0.033)&-0.028(-0.017)&0.004(-0.031)&-0.050(-0.068)&0.004(0.001)&-0.060(-0.047)\\
\hline
\end{tabular}
}
\caption{Top-quark charge asymmetries as a function of the rapidity ($y$) and the pseudo-rapidity ($\eta$), 
in \%, for the SM and the MSSM scenarios for $\beta_{z,t\bar{t}}>0$
($\beta_{z,t\bar{t}}>0.6$) at the $\sqrt{s}=14\TeV$ LHC.}
\label{tab:ACtt}
\end{center}
\end{table}

As widely argued in the literature, $t \bar t$ charge asymmetries could
appear in proton-proton collisions at the LHC (see for example~\cite{Aguilar-Saavedra:2013rza} and references therein).
In order to investigate them, 
we consider the cut-independent charge asymmetries defined by
Eq.~(\ref{eq:ACtt}), which are being used in current 
analysis by ATLAS~\cite{Aad:2013cea} and CMS~\cite{Chatrchyan:2014yta}.
 Both experiments are reporting charge 
asymmetries compatible with a zero value and consistent with the SM
expectations. In table~\ref{tab:ACtt} we present 
our results for the top-quark charge asymmetries, calculated from
Eq.~(\ref{eq:AC}), as a function of particle rapidity ($y$) 
and the pseudo-rapidity ($\eta$) at $\sqrt{s}=14 \TeV$. 
Results are given for the SM in $pp\to t\bar{t}$ channel and 
for all the four SUSY scenarios defined in the previous section.
It is important to remark here that the asymmetries are computed
for one model at a time, with the denominators at the same order as the 
numerators \footnote{The most precise current predictions for the cut-independent 
$t\bar t$ asymmetries
in the SM at the $14\TeV$ LHC are $A_{C}^{t\bar t}(y)=0.77\times10^{-2}$, 
$A_{C}^{t\bar t}(\eta)=0.59\times10^{-2}$~\cite{Kuhn:2011ri}. Our
results are not comparable with the previous ones.}.
Our SM values are computed
using a NLO approximation as included in {\tt MadGraph}~\cite{Alwall:2014hca}, 
they are meant for the comparison to the MSSM predictions.
At this point we want to emphasize that a detail understanding of the decay processes and 
experimental cuts in both models will be necessary to combine NLO level computations
in a consistent way. We only focused here on the impact of the effective 
corrections on top-quark charge asymmetries in several MSSM scenarios
and we compare their results with the SM predictions.
Our computations are done without any 
cut on the $z$-component of $t \bar t$-system velocity, $\beta_{z, t \bar t}$, 
and with a kinematic cut on $\beta_{z, t \bar t}$, 
requiring $\beta_{z, t \bar t} > 0.6$ (results within brackets of 
table~\ref{tab:ACtt}), which defines a region of phase space 
where the physics beyond the SM effects on the asymmetry may be 
enhanced~\cite{AguilarSaavedra:2011cp}. 
All the MSSM predictions, both at tree level and in 
the squark effective description approximation, are close to the SM ones,  
being the {\it{Def}} and \modmin scenarios the most compatible ones with the SM
predictions. As expected the $\beta_{z, t \bar t} > 0.6$ 
cut increases the values of the asymmetries in most of the cases.
The results for the top-quark charge asymmetries are also sensitive to the inclusion 
of radiative corrections through the analysis of the effective
approximations. For the case of $Def$ ($mod^-$)
scenario, when $\beta_{z, t \bar t} > 0.6$, the relative
deviation $\delta$ is about $-63\%$ ($43\%$) for $A_{C}^{t\bar{t}}(y)$,
$-53\%$ ($13\%$) for $A_{C}^{t\bar{t}}(\eta)$ and 
$-66\%$ ($-0.28\%$) for $A_{C}^{l^{+}l^{-}}(y)$.
Therefore, the top-quark asymmetries decrease strongly when radiative
corrections are included in the case of the $Def$ set but, contrary, in
the $mod^-$ scenario the situation changes a bit increasing the values
of $A_{C}^{t\bar{t}}$ whilst $A_{C}^{l^{+}l^{-}}(y)$ decreases.
In the other two scenarios, $pMSSM^c$ and {\it LS}, 
the results of the deviations are compatible with zero.
However, the radiative corrections can be also important,
being $\delta$ $-75$\% for $A_{C}^{t\bar{t}}(y)$ and 7\% for $A_{C}^{l^{+}l^{-}}(y)$
in the $pMSSM^c$ scenario, and around $-100$\% for $A_{C}^{l^{+}l^{-}}(y)$ in the {\it LS} set.
As a conclusion, the behavior of the top-quark asymmetries with the
inclusion of the radiative corrections is strongly dependent on the SUSY
scenarios, and it must be taken into account in phenomenological analysis.   

Let's emphasize that the $t\bar t$ charge asymmetry is a consequence 
of Feynman diagram interferences. Thus, it is necessary 
that at least two diagrams from the $q\bar q$ channel contribute significantly,
otherwise the asymmetry will be consistent with zero. 
Needless to say, all contributions from the $gg$ 
channel, despite they interfere with each other, are symmetric, and 
hence cancel when accounted in the calculation of Eq.~(\ref{eq:ACtt}).

First, we analyze the results for scenarios with the lowest values 
of the asymmetry: $pMSSM^c$ and $LS$.
In these scenarios, at tree level, more than 99\% of the $q\bar q$
contributions to the total cross section comes from
the double resonant diagrams as in figure~\ref{fig:pairdiag}{\bf{a}} 
with $\squark\to\stopp_{1}$. 
When using the effective approximation of squark interactions, 
the relative weight of $q\bar q$ diagrams 
do not change and the mentioned diagram is by far kept as the leading one. 
Then, for these scenarios there is no chance of observing 
values of $A_{C}^{t\bar t}$ differing 
from zero regardless the approximation used.

Second, the $\modmin$ scenario gives a positive asymmetry a bit larger 
than in the above two scenarios.
In this case, the diagrams as in figure~\ref{fig:pairdiag}{\bf{a}} with $\squark\to\stopp_{1}$
give a contribution of $43\% \,(48\%)$ to the $q\bar q$ cross section
at tree level (in the effective approximation). The contribution with $\squark\to\stopp_{2}$ 
is less than $1.5\%$. Correspondingly, 
a $45\% \,(42\%)$ of the $q\bar q$ contribution is absorbed by diagrams as
in figure~\ref{fig:pairdiag}{\bf{b}}, being $\tilde{q}\to\stopp_{1}$ and
$\tilde{f}\to\tilde{f}_{1,2}$ the squarks of the first and the second
generation. These squarks in the $\modmin$ scenario
are degenerate and the physical masses are around
$1.5\TeV$ thus enhancing the interference
and, therefore, changing $A_{C}^{t\bar t}$ towards values different from zero.
In this scenario, the interference between $\stopp_{1}$ and $\stopp_{2}$
diagrams is small. Summarizing, the main 
interference comes from diagrams like in Figure~\ref{fig:pairdiag}{\bf{b}}
with the first and second generation of squarks in the internal line 
and the lightest top-squark.

Finally, the $Def$ scenario also has a different behavior of the relative weights of 
the contributions to the $q\bar q$ cross section.
Diagrams of figure~\ref{fig:pairdiag}{\bf{a}} with
$\tilde{q}\to\stopp_{1,2}$ give a 75\% (81\%)
of the total cross section of the $q\bar q$ channel and split on 43\% (55\%) for the $\stopp_{1}$
and 32\% (26\%) for the $\stopp_{2}$ diagram at tree level (in the effective calculation).
In this scenario the larger branching ratio of the decay 
$\stopp_{2}\to t\tilde \chi_1^0$ in comparison to the branching ratio of the 
$\stopp_{1}\to t\tilde \chi_1^0$ compensates the difference of each pair
production cross section, allowing
the interference between each other.
Diagrams as in figure~\ref{fig:pairdiag}{\bf{b}} with
$\tilde{f}\to\tilde{f}_{1}$ in the internal line
contribute in about 15\% (12\%) to the $q\bar q$ channel cross section, 
evenly split between $\stopp_{1}$ and $\stopp_{2}$
and hence interfering between each other as well. 
The diagrams with $\tilde{f}\to\tilde{f}_{2}$ contributes less than a $0.5\%$. 
Therefore, the asymmetry values arises 
from the interference between the diagrams
as in figure~\ref{fig:pairdiag}{\bf{a}} and~\ref{fig:pairdiag}{\bf{b}} 
with both squarks, $\stopp_{1}$ and $\stopp_{2}$, 
and the lightest squark of the first and second generation.
In summary, the reasons to explain asymmetry values different 
from zero in the MSSM
are rather dependent on the SUSY parametrization chosen.

\begin{table}[t]
\begin{center}
\begin{tabular}{|c|cccccc|}
\hline
&\multicolumn{6}{c|}{$A_{t\bar{t}}(Y_\text{cut}>0.7)\times 100$}\\
&\multicolumn{2}{c}{$m_{t\bar{t}}>2m_{t}$}&\multicolumn{2}{c}{$m_{t\bar{t}}<450\GeV$}&\multicolumn{2}{c|}{$m_{t\bar{t}}>450\GeV$}\\
\hline
SM $pp\to t\bar{t}$&\multicolumn{2}{c}{0.304}&\multicolumn{2}{c}{0.301}&\multicolumn{2}{c|}{0.424}\\
\hline
$pp\to t\bar{t}\tilde \chi_1^0\tilde \chi_1^0$&Tree&Eff&Tree&Eff&Tree&Eff\\
\hline
{\it Def}&0.806&0.475&0.711&-0.195&0.792&0.509\\
\modmin&0.339&0.290&-0.607&-0.422&0.387&0.331\\
\pMSSM&-0.011&0.077&0.295&-0.125&-0.055&0.081\\
LS&0.048&-0.039&-0.126&-0.144&0.079&0.048\\
\hline
\end{tabular}
\caption{Top-quark charge asymmetries $A_{t\bar{t}}(Y_\text{cut})$, in \%, 
for $Y_\text{cut}>0.7$ when $m_{t\bar{t}}<450\GeV$ and
$m_{t\bar{t}}>450\GeV$ for the SM in $pp\to t\bar{t}$ 
and MSSM scenarios at the $\sqrt{s}=14\TeV$ LHC.}
\label{tab:topasymmetriesymean}
\end{center}
\end{table}
Since most of the charge asymmetries are focused on large
rapidities~\cite{Kuhn:2011ri}, we also use the 
complementary definition of asymmetry on Eq.~(\ref{eq:ACymean}) in our
analysis. Table~\ref{tab:topasymmetriesymean} summarizes 
the results of $A_{t \bar t}(Y_\text{cut})$ for the SM and our SUSY 
scenarios with $Y_\text{cut} = 0.7$ and different cuts 
on $m_{t \bar t}$\footnote{The current most precise prediction for 
$A_{t\bar t}(Y_{\text{cut}}=0.7)$ at the $14\TeV$ LHC are: $0.75\times10^{-2}$ ($m_{t\bar t}<450\GeV$) and
$1.21\times10^{-2}$ ($m_{t\bar t}>450\GeV$)~\cite{Kuhn:2011ri}.}. 
Recall that, by definition, 
$A_C^{t \bar t}(y)|_{\beta_{z, t \bar t} > 0} = A_{t \bar t} \,(Y_\text{cut} = 0)$. 
We see clearly that the 
SM predictions are very dependent on the $m_{t \bar t}$ cut, 
which enhances almost a factor of 1.5 the value of the asymmetry 
from $m_{t \bar t} > 2 m_t$ to $m_{t \bar t} > 450$ GeV. 
$LS$ and $pMSSM^c$ predictions also increase with the $m_{t \bar t}$
cut, while $Def$ and $\modmin$ ones hardly change. 
The MSSM scenario most compatible with the SM predictions is 
$\modmin$, while $Def$ present results of the same order of magnitude but no so close to
the SM ones. Clearly, for $\pMSSM$ and {\it LS} scenarios the results are at least one order 
of magnitude smaller, but also consistent within the statistical uncertainties.

\begin{figure}[t]
\centering
\resizebox{\textwidth}{!}{
\begin{tabular}{ccc}
\includegraphics[width=6.4cm,angle=0]{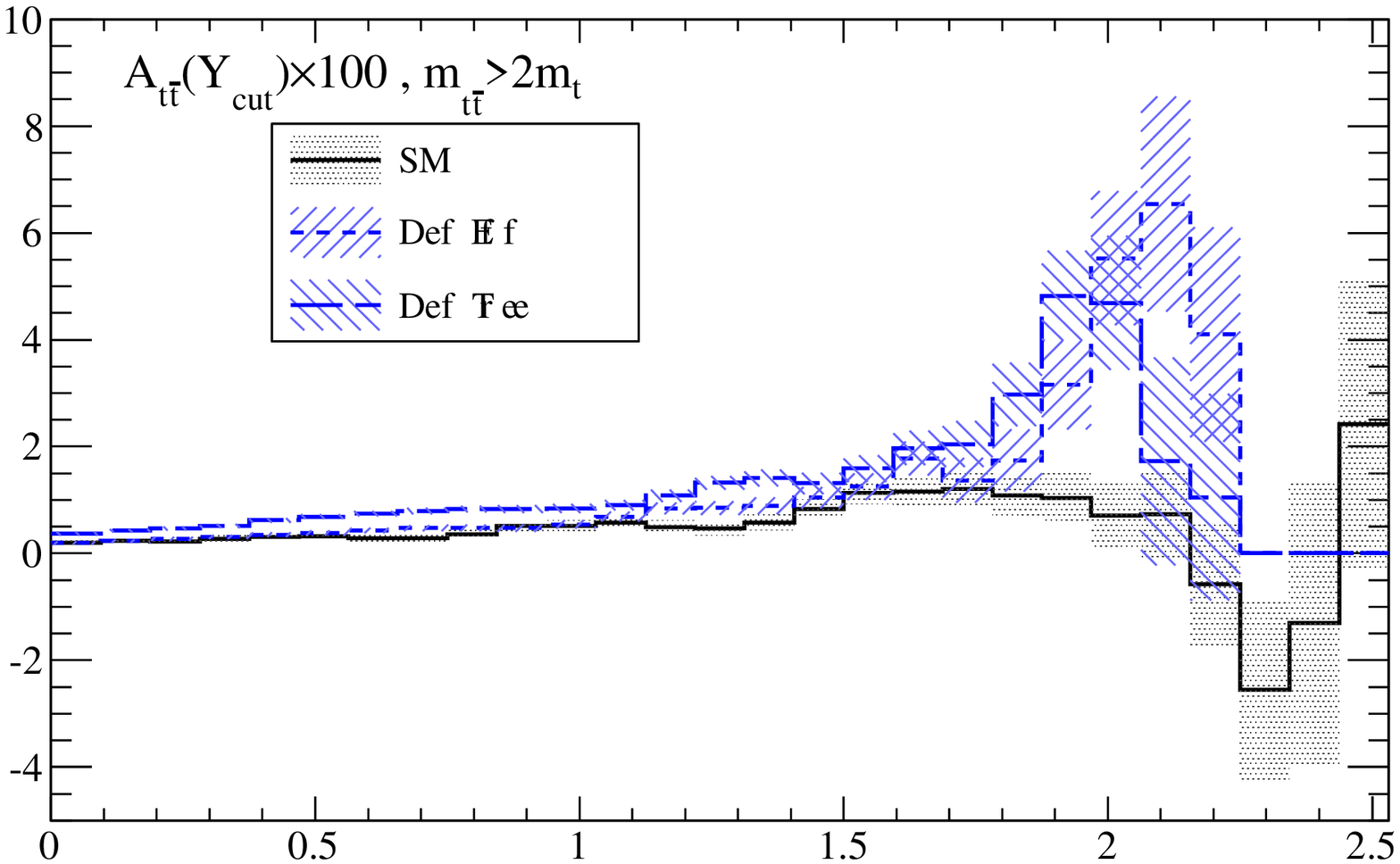}&
\includegraphics[width=6.4cm,angle=0]{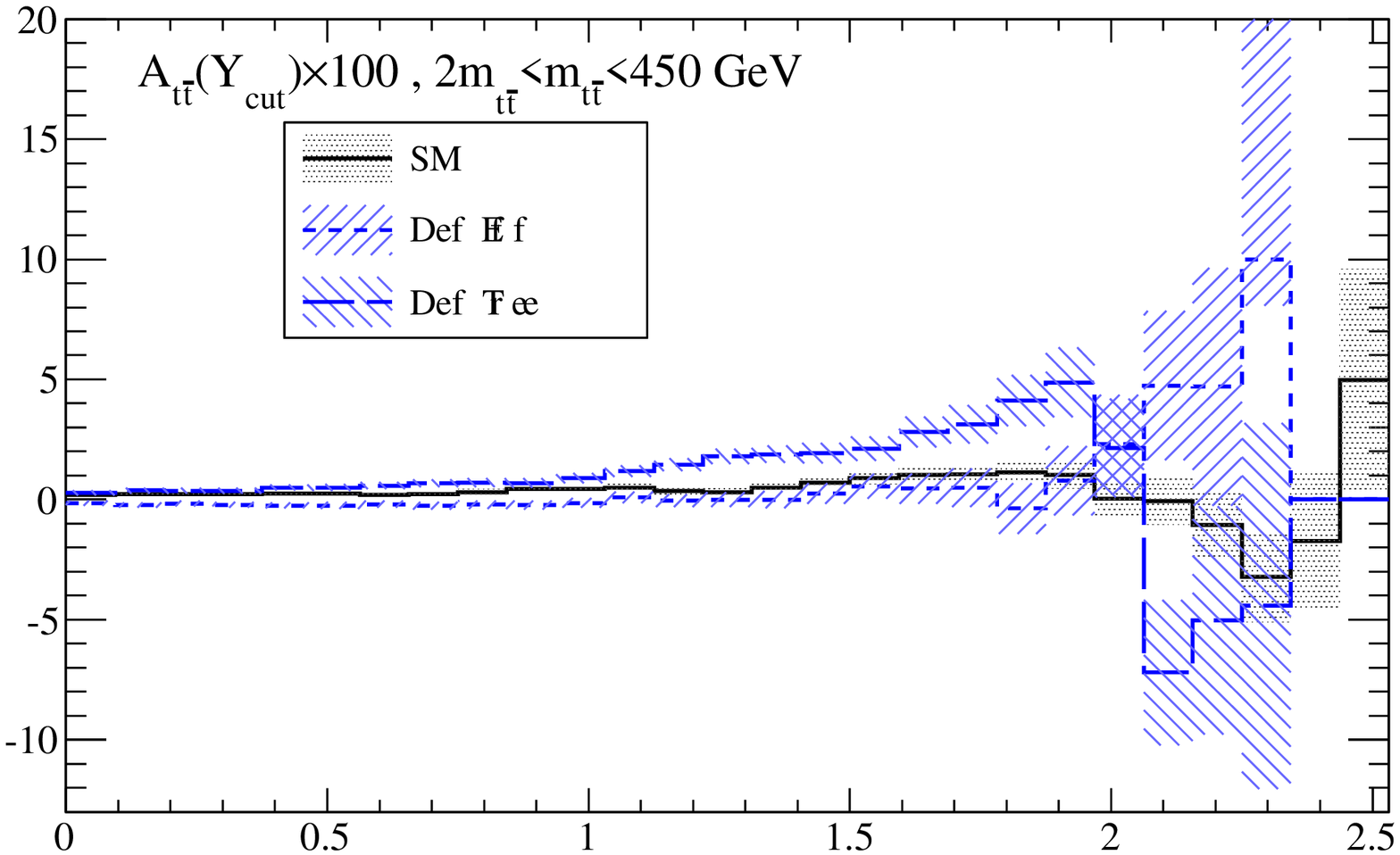}&
\includegraphics[width=6.4cm,angle=0]{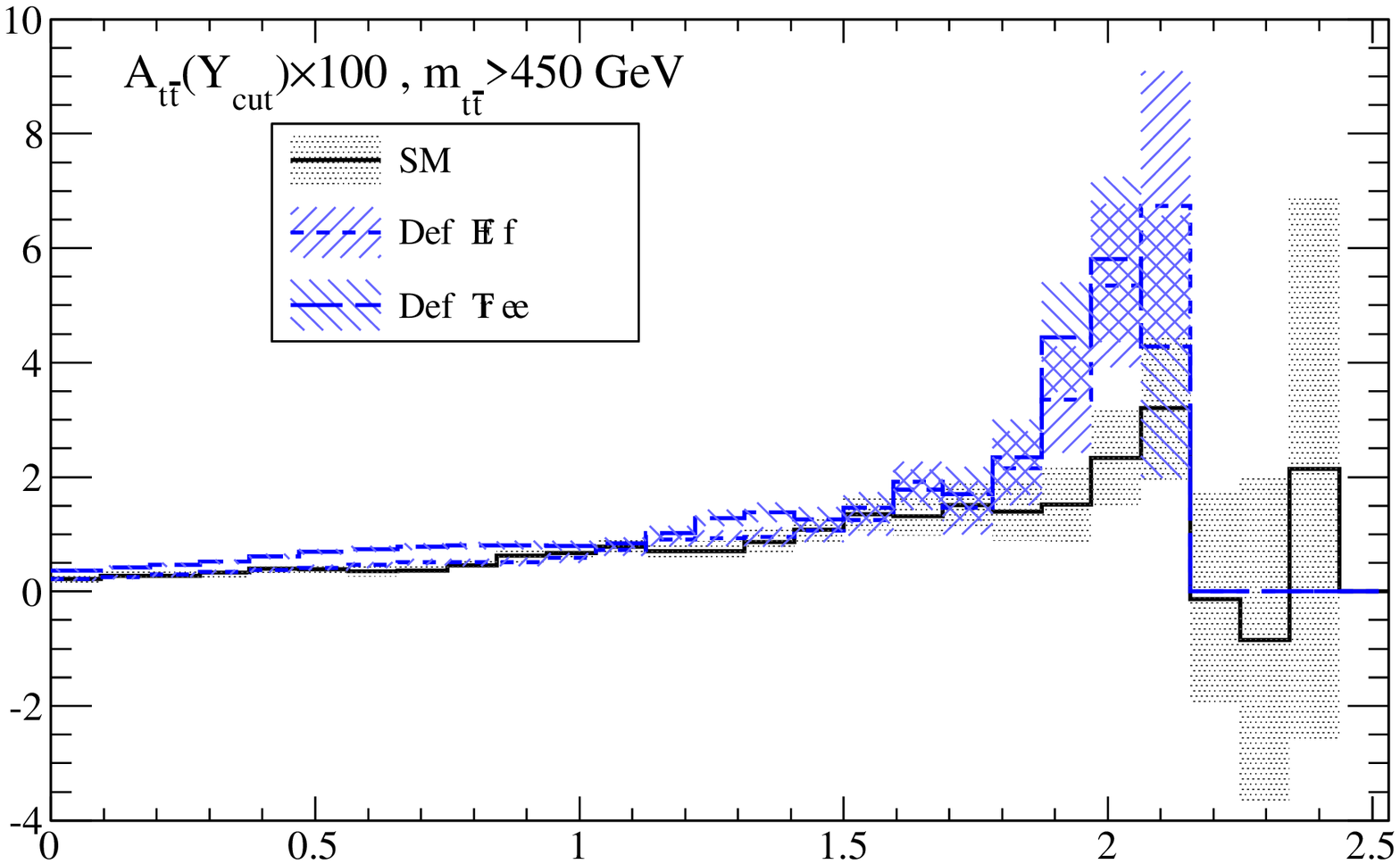}\\
\includegraphics[width=6.4cm,angle=0]{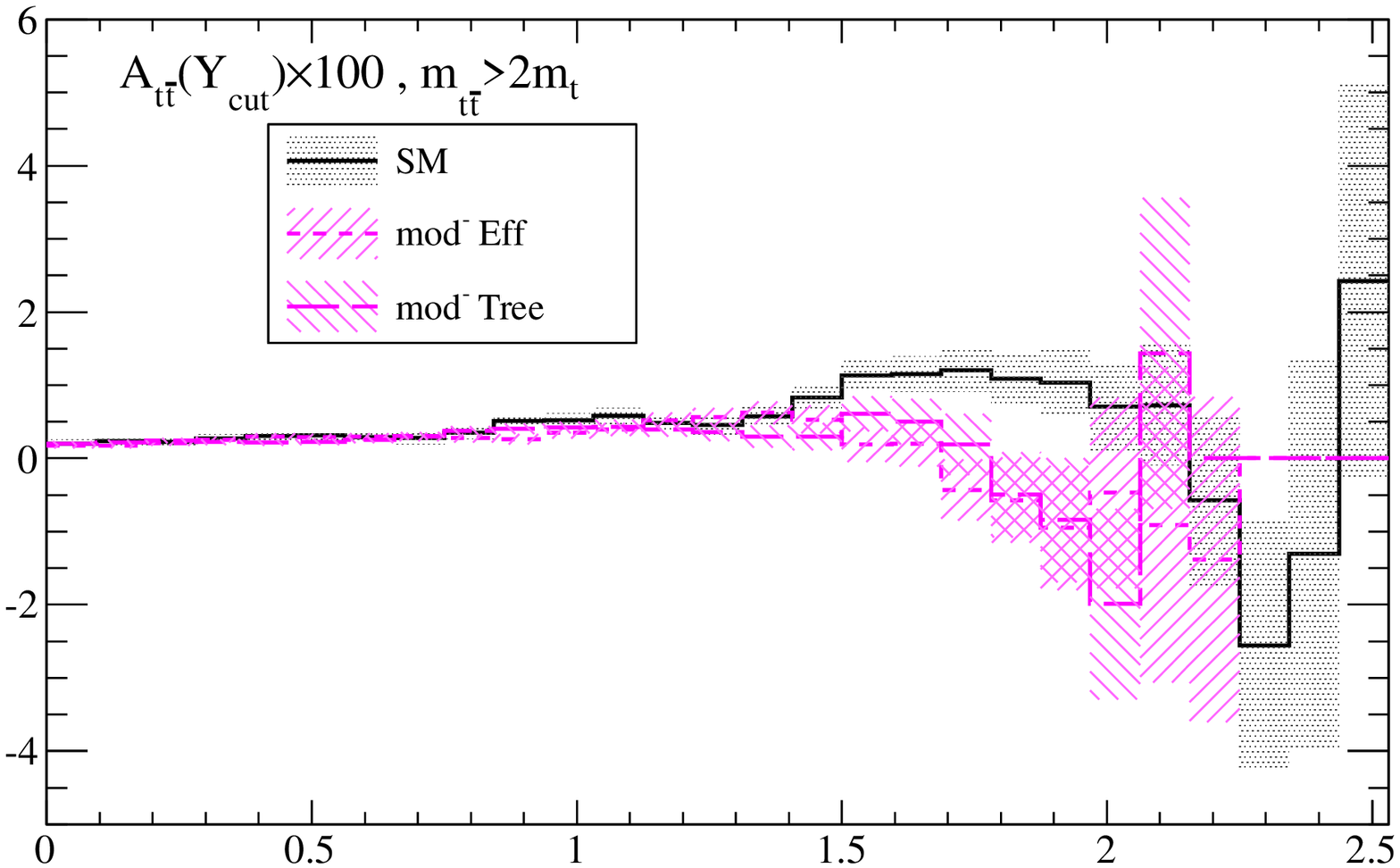}&
\includegraphics[width=6.4cm,angle=0]{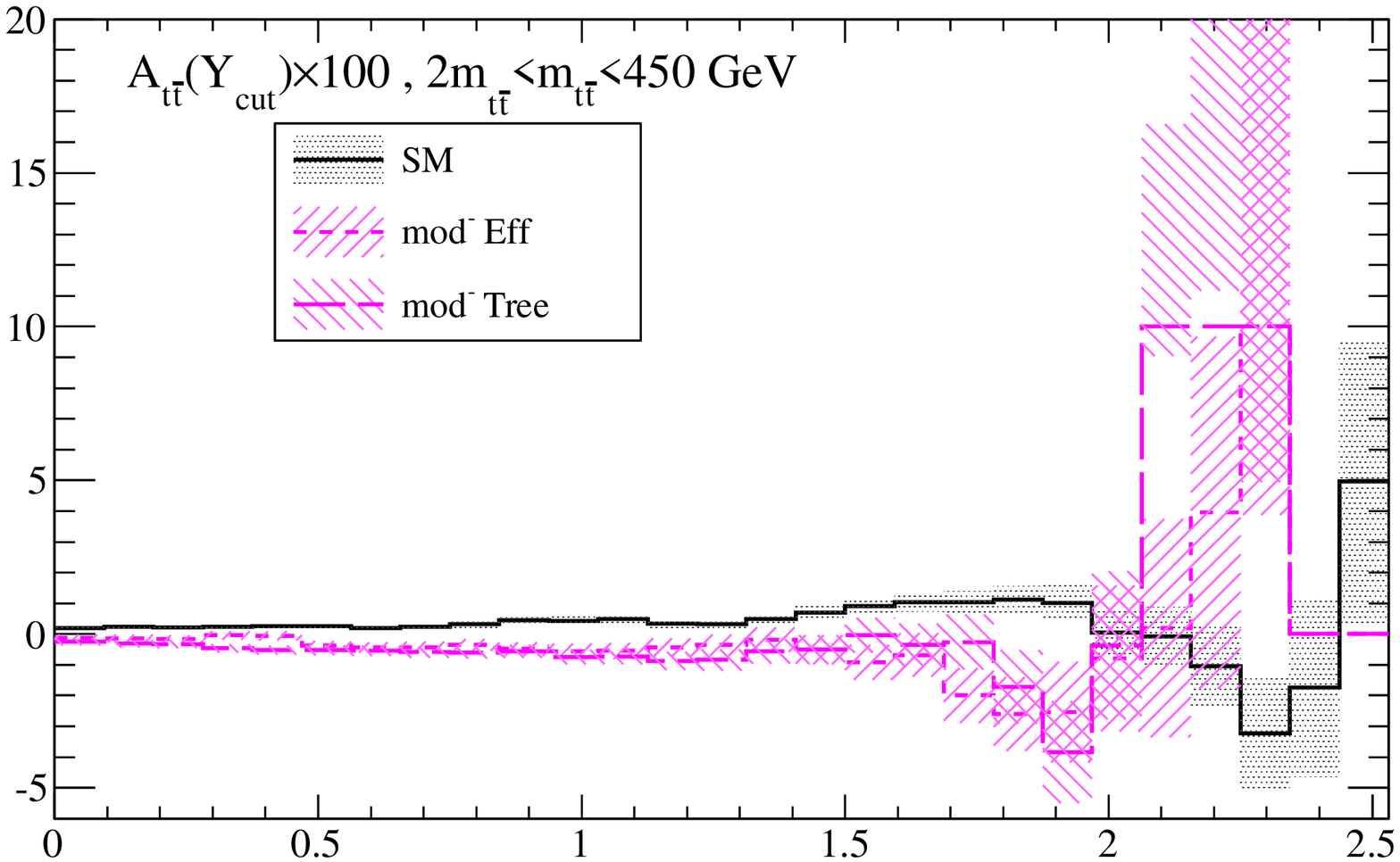}&
\includegraphics[width=6.4cm,angle=0]{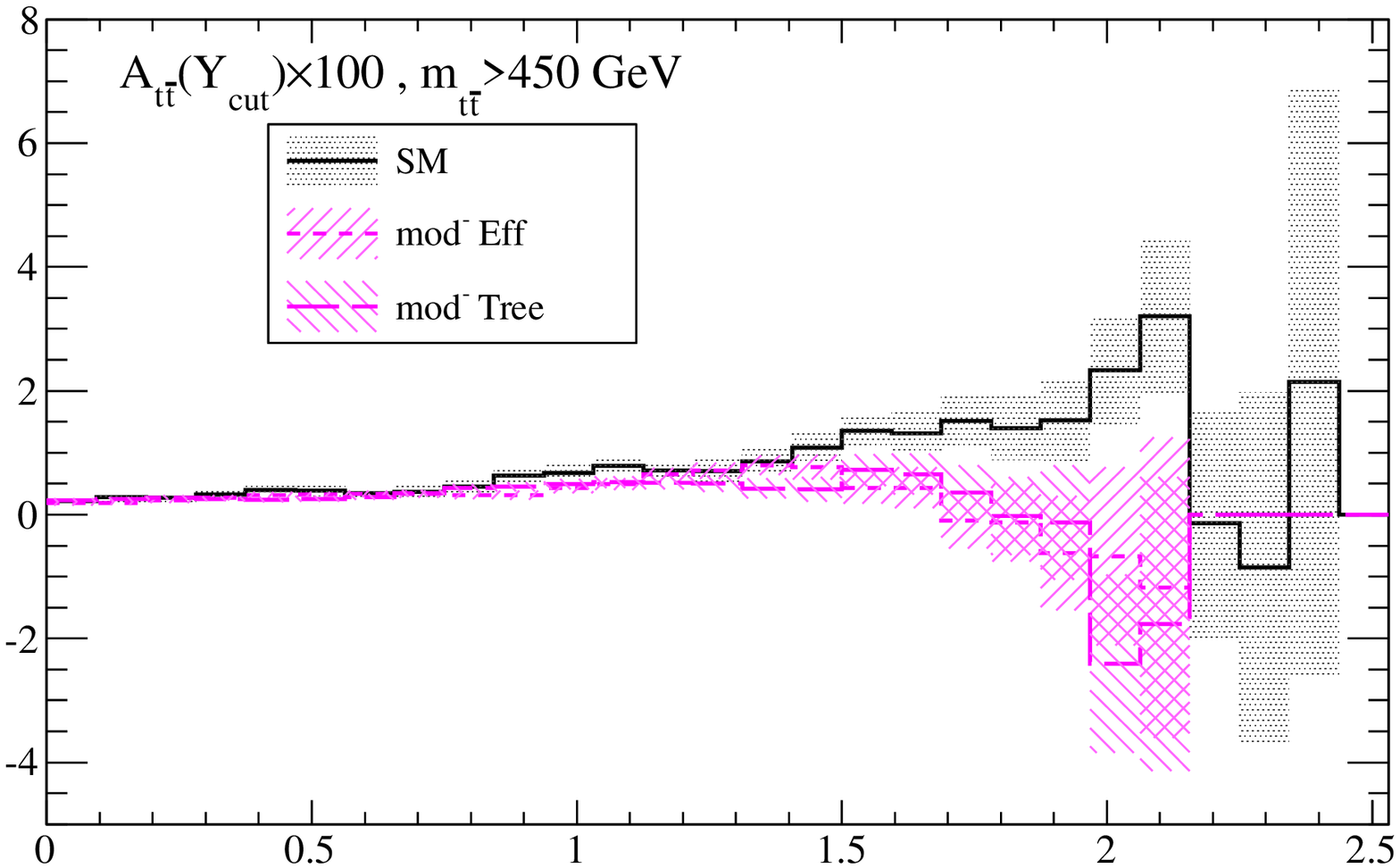}\\
\includegraphics[width=6.4cm,angle=0]{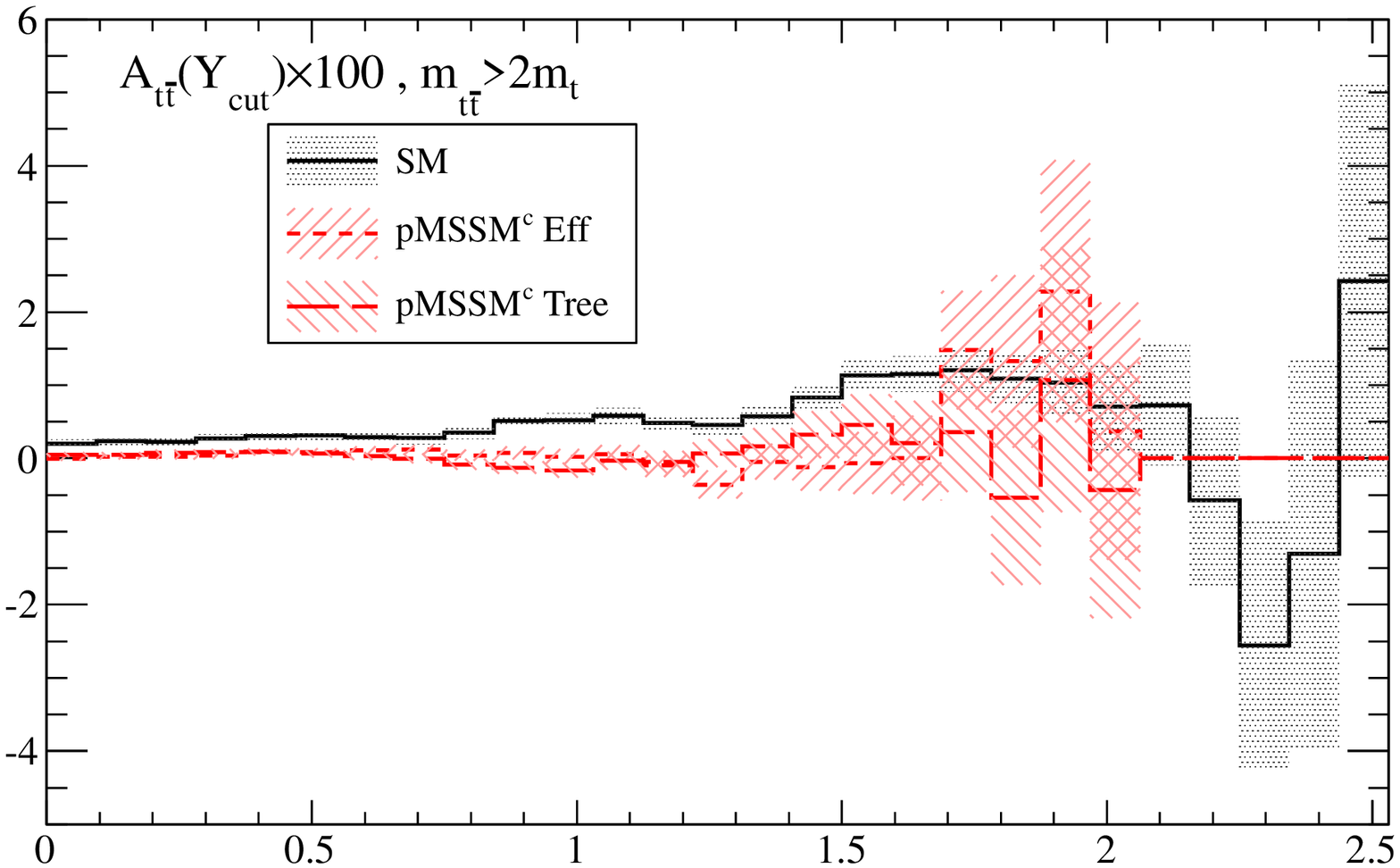}&
\includegraphics[width=6.4cm,angle=0]{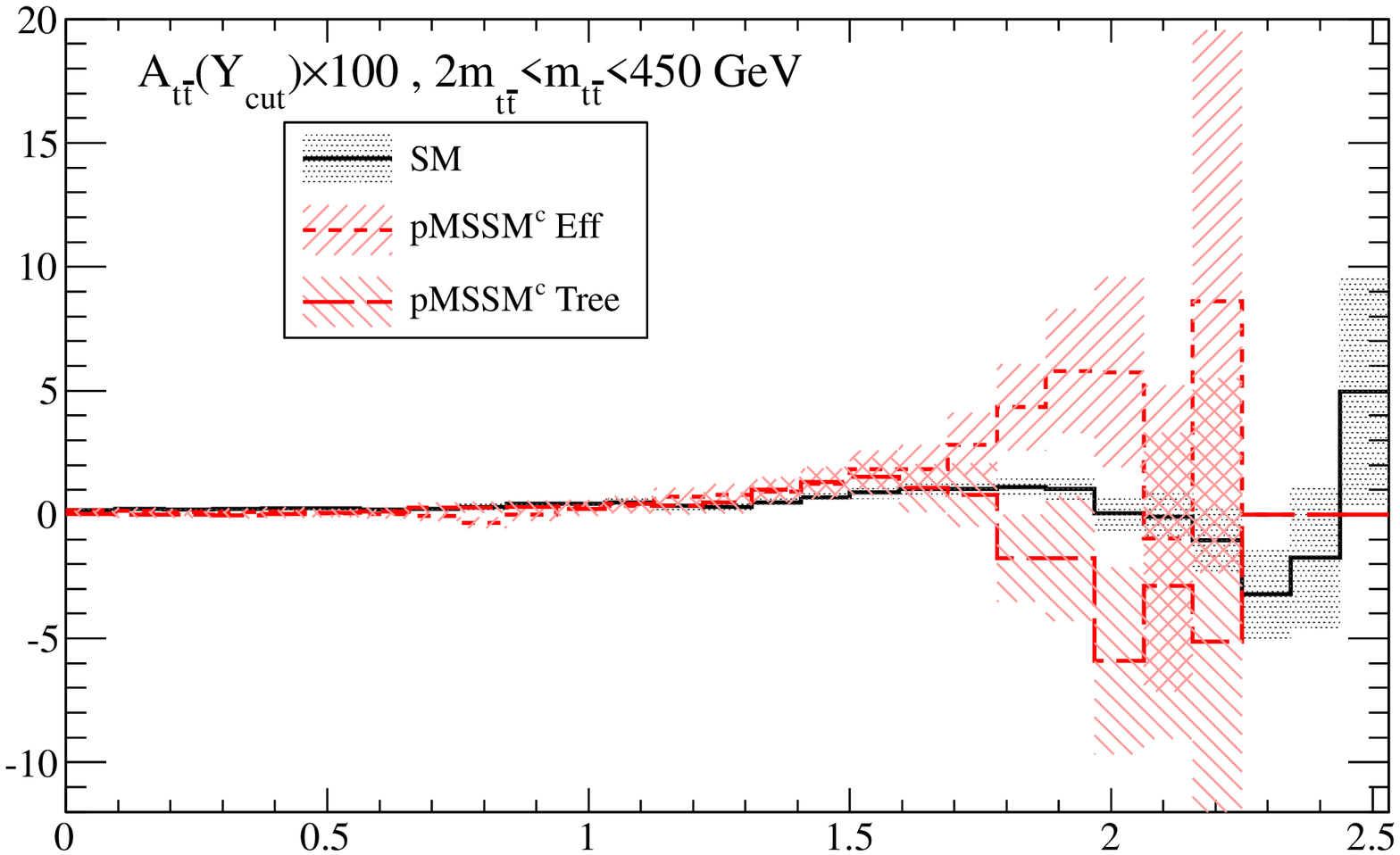}&
\includegraphics[width=6.4cm,angle=0]{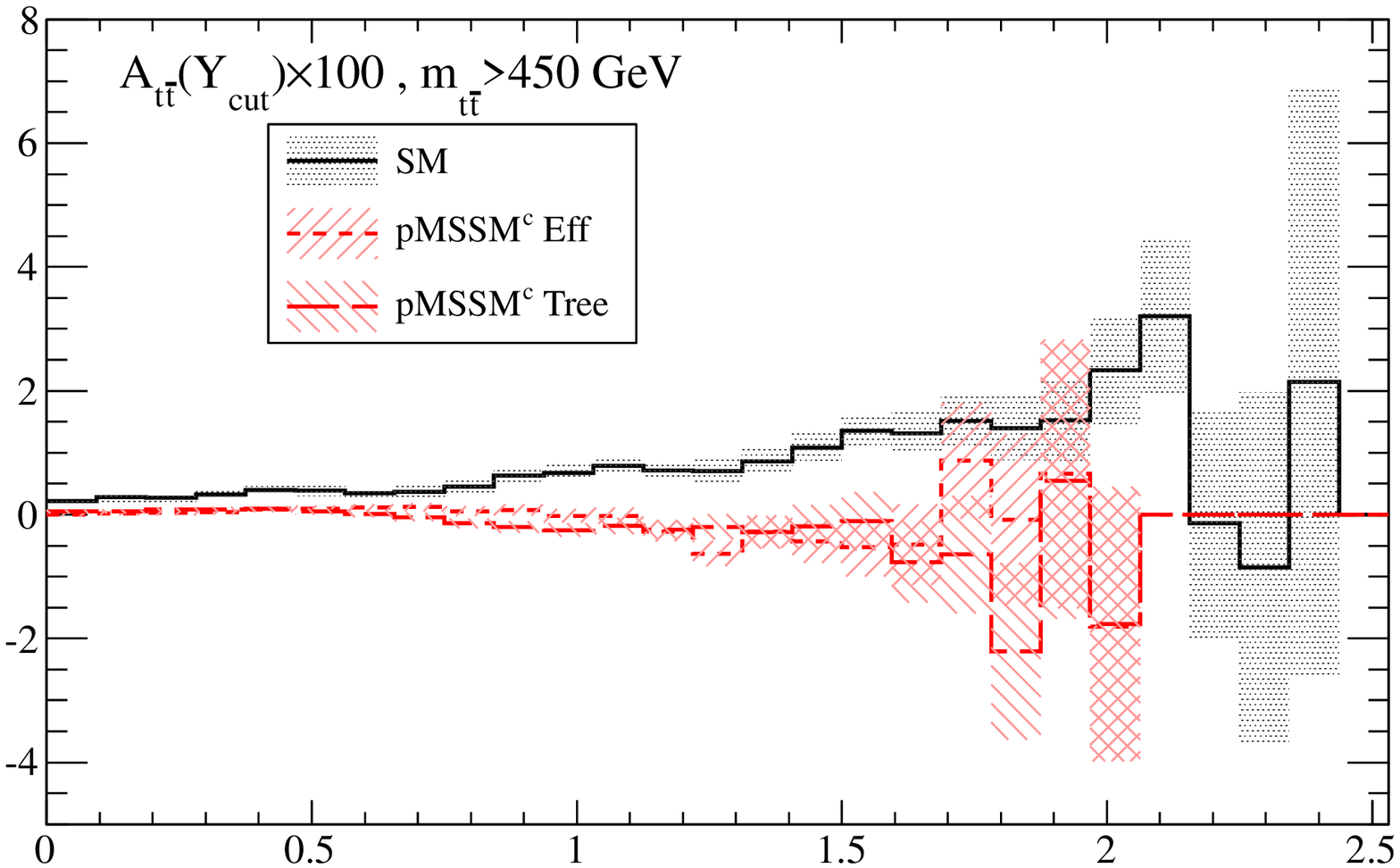}\\
\includegraphics[width=6.4cm,angle=0]{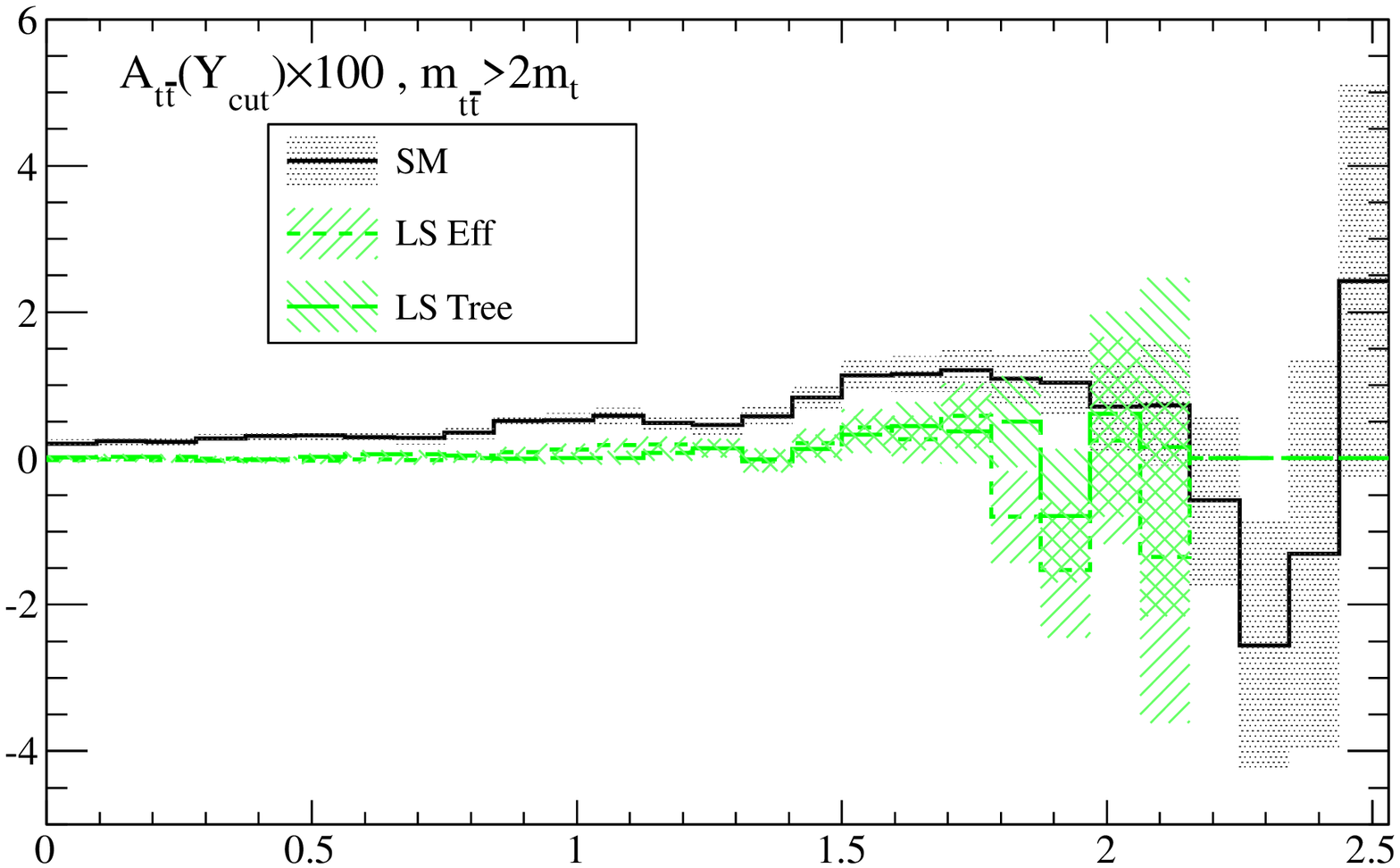}&
\includegraphics[width=6.4cm,angle=0]{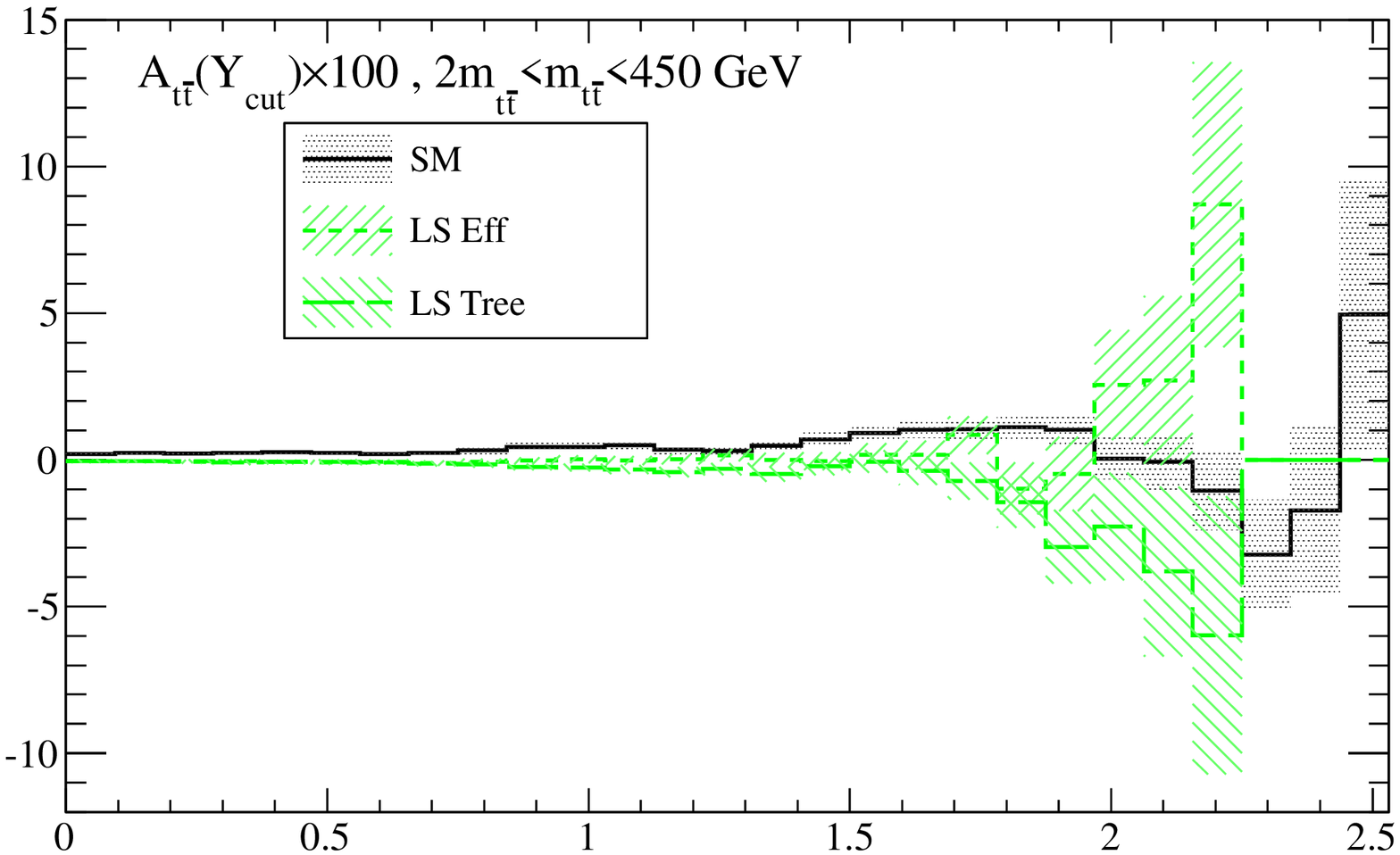}&
\includegraphics[width=6.4cm,angle=0]{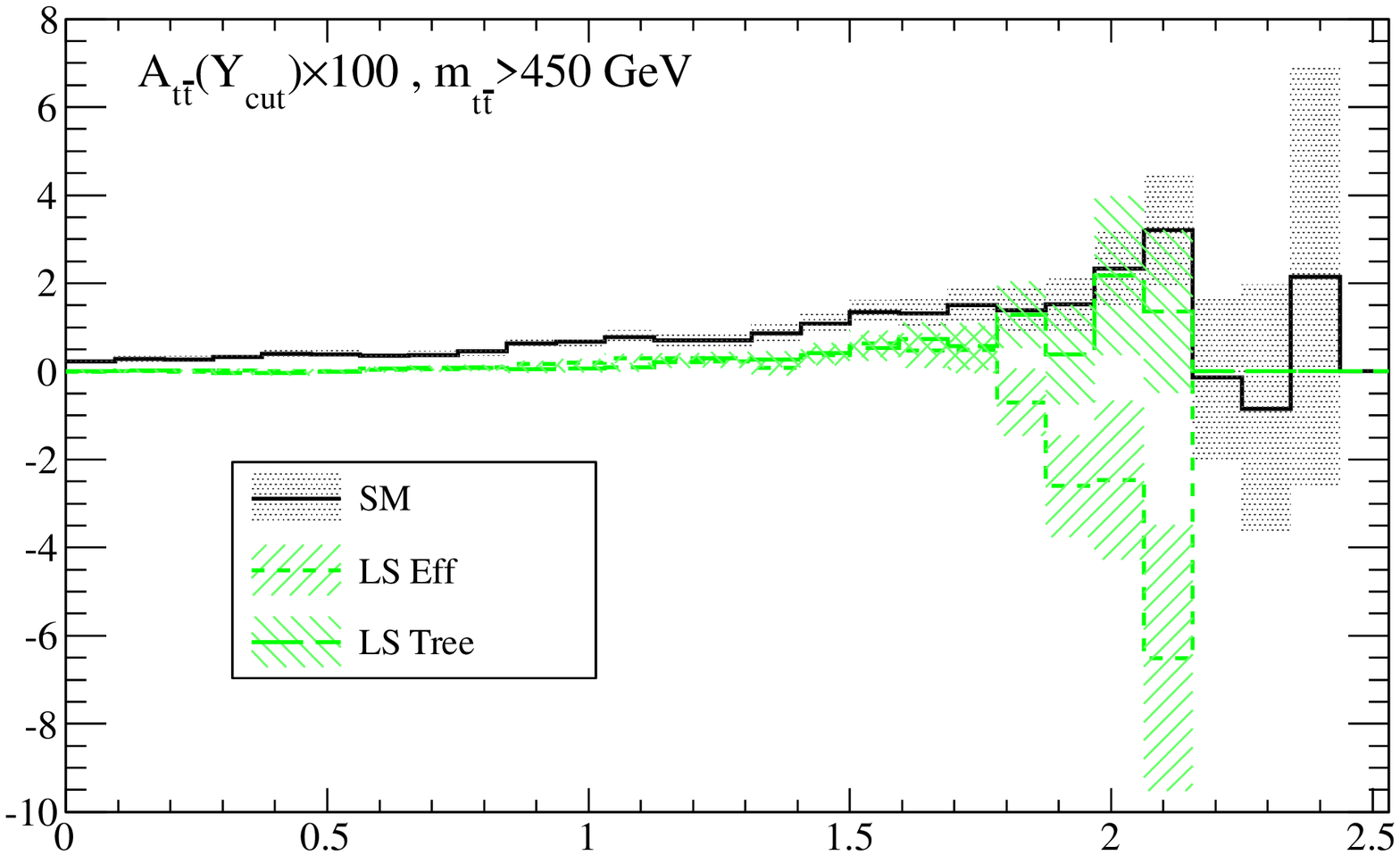}\\
(a)&(b)&(c)\\
\end{tabular}
}
\caption{$A_{t\bar{t}}(Y)$ vs. $Y_\text{cut}$ for the SM and all SUSY scenarios:
~{\bf (a)} in the whole range of $m_{t\bar{t}}$,
~{\bf (b)} for $m_{t\bar{t}}<450\GeV$ and ~{\bf (c)} 
for $m_{t\bar{t}}>450\GeV$ at the $\sqrt{s}=14\TeV$ LHC.}
\label{fig:ACYmeanallScen}
\end{figure}

In order to find kinematic regions where the top-squark-pair production via 
$q \bar q$ annihilation was comparable or 
even dominant over gluon fusion, and thus the charge asymmetry is
capable to help in the search for SUSY, we show in 
figure~\ref{fig:ACYmeanallScen} the behavior of the asymmetry 
$A_{t \bar t}(Y_\text{cut})$ for all the four SUSY 
scenarios in comparison with the SM predictions, and for different 
values of the $m_{t \bar t}$ cut. 
Each bin of this figure contains at least 0.01$\%$ 
of the total number of simulated events (500 events). 
Large values of $Y_{\text{cut}}$ contain fewer events and are not shown. 
With this set up the largest statistical uncertainty is 4.5\%.  
We can conclude the following statements for each scenario:
\begin{itemize}
\item {\it Def}: The tree and the effective approximation results 
are very similar in this scenario and both 
predictions are almost equal to the SM ones, specially for low values of 
$Y_\text{cut}$. The $m_{t\bar t}>450\GeV$ cut shows similar results as 
the non-cut asymmetry.
\item $\modmin$: In this scenario there is no difference between $m_{t \bar t} > 2 m_t$ and $m_{t \bar t} > 450$ GeV cuts. 
In both cases and for low values of $Y_\text{cut}$, the predicted asymmetry is close to the SM ones. 
For $Y_\text{cut} >$ 1.5, the asymmetry becomes negative and could be distinguishable from the SM prediction. 
Again, the results with $m_{t \bar t} < 450$ GeV are very different to the other two cuts, and we obtain 
very different results for the tree and the effective approximations, 
with the predictions of the latter very similar to the SM results.
\item $\pMSSM$: All the predictions in this scenario, in both tree and 
effective approximations and for low values of $Y_\text{cut}$,
are compatible with no asymmetries. The best region in order to try to distinguish 
this scenario from the SM is $1.5 < Y_\text{cut} < 2.2$, 
with $m_{t \bar t} < 450$ GeV, 
in which the effective description provides a much larger asymmetry 
than the SM one whilst the tree-level results are also large but negative.
\item  $LS$: For low values of $Y_\text{cut}$, this scenario is also compatible with no asymmetries, while in the 
effective approximation and for large values of $Y_\text{cut}$ the charge asymmetry becomes negative
for $m_{t \bar t} > 2 m_t$ and $m_{t \bar t} > 450$ GeV. 
The use of a cut in $m_{t \bar t}$ also enhances the size of the asymmetry 
for $Y_\text{cut}$ larger than $2$. On the other hand, the behavior of the asymmetry 
is very different if we consider the tree-level results. 
In this case, the $m_{t \bar t}$ cut does not help and it is very difficult to differentiate the tree-level results of 
$LS$ from the SM predictions.
\end{itemize}

The main conclusion of this section is that all the SUSY predictions 
of $A_\text{C}$ are compatible with the SM ones.
For low values of $Y_\text{cut}$, it is hard to distinguish between the
SM and the SUSY results. 
On the other hand, for values of
$Y_\text{cut} >$ 1.5 the MSSM predictions are very different to the SM ones. 
However, the statistical uncertainties in these cases
are so large that do not allow us to draw any conclusion. 
In other words, we cannot make use of 
these $A_\text{C}$ results in order to discriminate
among the SUSY scenarios proposed along this work. 
Fortunately, the study of top-quark 
polarizations may provide additional information,
useful to differentiate among scenarios as we will see in the next section.

\subsection{Top-quark polarization}

\begin{table}[t]
\begin{center}
\begin{tabular}{|c|cc|cc|cc|}
\hline
$\mathcal{P}_{t}$&\multicolumn{2}{c|}{$\tb=10$}&\multicolumn{2}{c|}{$\tb=30$}&\multicolumn{2}{c|}{$\tb=50$}\\
$\stopp_1 \to t \tilde \chi_1^0$&Tree&Eff&Tree&Eff&Tree&Eff\\
\hline
{\it Def}&0.17&0.39&0.27&0.44&0.29&0.44\\
\modmin&0.83&0.87&0.89&0.93&0.90&0.94\\
\pMSSM&-0.98&-0.95&-0.96&-0.93&-0.96&-0.92\\
{\it LS}&0.32&0.44&0.43&0.52&0.44&0.53\\
\hline
\end{tabular}
\caption{Polarization of top quarks, $\mathcal{P}_{t}$, coming from 
$\stopp_1 \to t\tilde \chi_1^0$ decays for the MSSM scenarios with $\tb=10,30,50$.}
\label{tab:polarizationResume}
\end{center}
\end{table}

\begin{table}[h]
\begin{center}
\begin{tabular}{|c|cc|cc|cc|}
\hline
&\multicolumn{2}{c|}{$A_{\phi_{l}}$}&\multicolumn{2}{c|}{$A_{\theta_{l}}$}&
\multicolumn{2}{c|}{$\mathcal{P}_{t}$}\\
\hline
SM $pp\to
t\bar{t}$&\multicolumn{2}{c|}{0.6165}&\multicolumn{2}{c|}{0.3134}&
\multicolumn{2}{c|}{0.0}\\
\hline
$pp\to t\bar{t}\tilde \chi_1^0\tilde \chi_1^0$&Tree&Eff&Tree&Eff&Tree&Eff\\
\hline
{\it Def}&0.9483&0.9522&0.8214&0.8301&0.17&0.39\\
\modmin&0.9689&0.9713&0.8824&0.8903&0.87&0.91\\
\pMSSM&0.9381&0.9390&0.7174&0.7214&-0.96&-0.93\\
LS&0.8514&0.8626&0.5636&0.5892&0.32&0.44\\
\hline
\end{tabular}
\caption{Lepton asymmetries for the SM in $pp\to t\bar{t}$ -first row-
  and in the MSSM in $pp\to t\bar{t}\tilde \chi_1^0\tilde \chi_1^0$
at the $\sqrt{s}=14\TeV$ LHC.}
\label{tab:leptonasymmetries}
\end{center}
\end{table}
The longitudinal polarization of the top quarks coming from top-squark decays 
into neutralinos may supply information about the SUSY scenario under 
study~\cite{Godbole:2006tq}. As it is known, it also differs from the
unpolarized SM pair-produced top quarks. We investigate in this work if 
the inclusion of the radiative corrections to quark-squark-gaugino 
couplings~\cite{Guasch:2008fs} may change the final 
polarization state of the top-quarks in the SUSY framework.

First of all, we show in table~\ref{tab:polarizationResume} 
the expected longitudinal top-quark polarization for each SUSY scenario, 
calculated with Eq.~(\ref{eq:polarization}). As explained in~\cite{Belanger:2012tm,Belanger:2013gha}, 
the value and the sign of polarization of tops coming from 
$\tilde t_{1,2} \to t \,\tilde \chi_1^0$ decays depend on the mixing of the
top-squark sector, the nature of the neutralino and the mass difference between the top-squark and the neutralino. 
In all the SUSY scenarios evaluated in this work, $\tilde \chi_1^0$ is a pure bino-like neutralino. 
Such bino-like neutralino couples stronger to the right-handed (RH) than 
to the left-handed (LH) components of the top-squark, 
enhancing positive values of the polarization even though the LH component of the top-squark was larger. Anyhow, 
whenever the LH (RH) component of the top-squark is overwhelmingly above the RH (LH) one, and the $f_{2}$ 
factor in Eq.~(\ref{eq:polarization}) vanishes\footnote{This factor does
  not trivially vanishes for large mass difference between the top-squark and its decays products. It also
needs a relatively small mass for the gaugino to be negligible.}, the polarization will have values of $\mathcal{P}_{t}=-1$ $(+1)$.

In the $pMSSM^c$ scenario, the lightest top-squark $\tilde t_1$ is mostly LH and therefore we obtain a top polarization very close to 
$-1$, as shown in table~\ref{tab:polarizationResume}.
The $mod^-$ scenario represents a parametrization where the mixing in the top-squark sector is maximal
and the large mass difference between the top-squark and the decay products induces $f_{2}\to0$. 
Due to the fact that the bino-like neutralino
couples stronger to the top-squark RH component than to the LH one and $f_{2}\to0$, we obtain
large values of the polarization, $\sim0.8$, in agreement with results in~\cite{Belanger:2012tm} (see figure 3 therein).
In $Def$ and {\it LS} scenarios the top-squark mixing has similar and large LH components but with an important RH components
which contribute more to the top-quark polarization than the LH ones.
In these scenarios the values of the LH and RH couplings (which are of the same sign) 
in Eq.~(\ref{eq:polarization}) are rather similar.
The increase of the polarization values for the {\it LS} scenario with respect to $Def$ scenario
lies on the difference among the value of $f_{2}$ in Eq.~(\ref{eq:polarization}).
In the $Def$ scenario $f_{2}\to 0$ meanwhile in the {\it LS} scenario $f_{2}\approx0.5$.
The denominator of Eq.~(\ref{eq:polarization}) for the {\it LS} scenario is smaller
than for the $Def$ scenario and then the polarization is larger for the former than for the latter.
A small variation on the values of $\mathcal{P}_{t}$ can be also
appreciated as a function of $\tan\beta$, since this parameter slightly
modifies both the top-squark and neutralino mixings, and consequently the LH
and RH components of top-squarks and the nature of the lightest
neutralino, $\tilde \chi_1^0$.

\begin{figure}[t]
\centering
\resizebox{\textwidth}{!}{
\begin{tabular}{cc}
\includegraphics[width=6.0cm,angle=0]{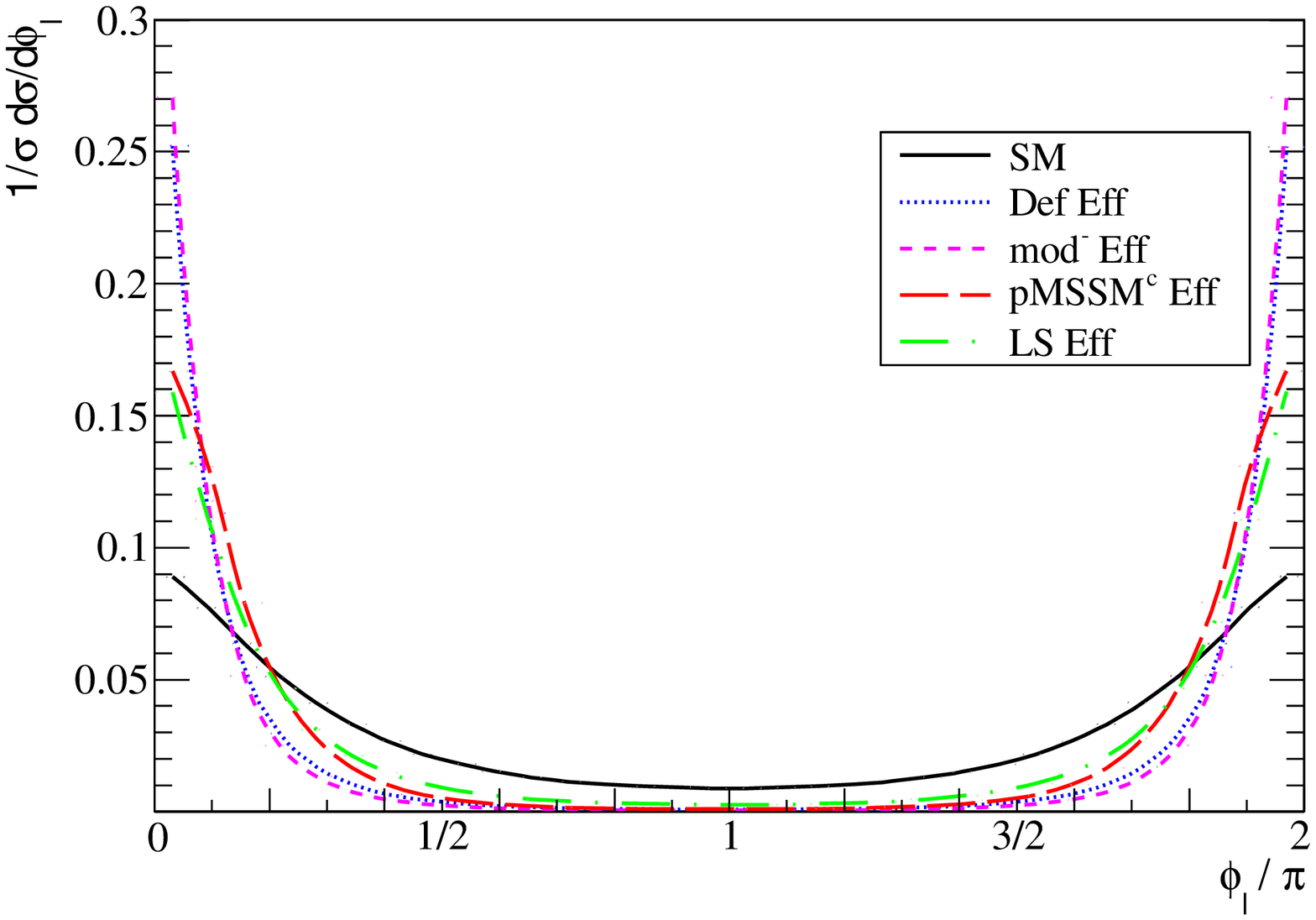}&
\includegraphics[width=6.0cm,angle=0]{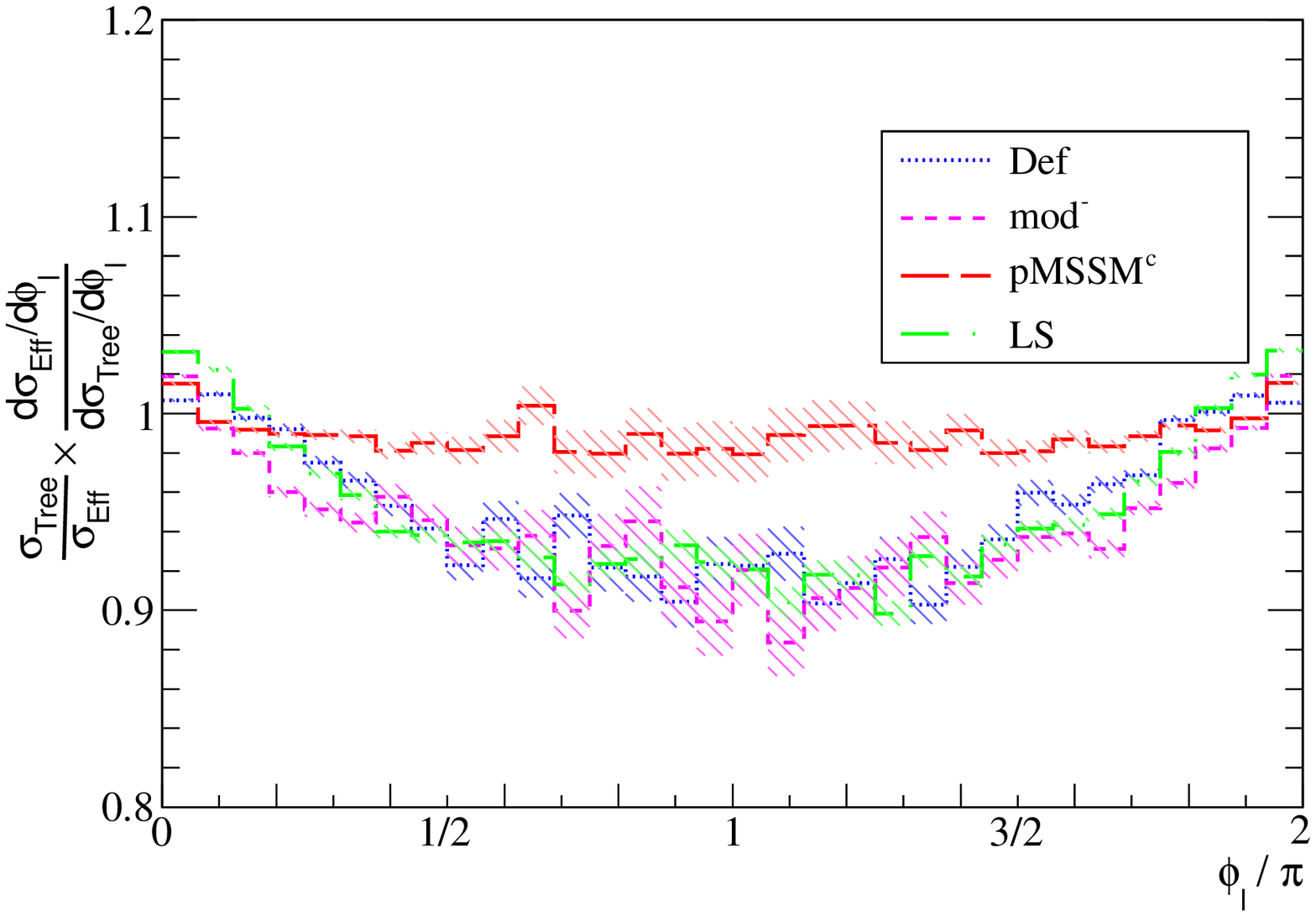}\\
\multicolumn{2}{c}{(a)}\\
\includegraphics[width=6.0cm,angle=0]{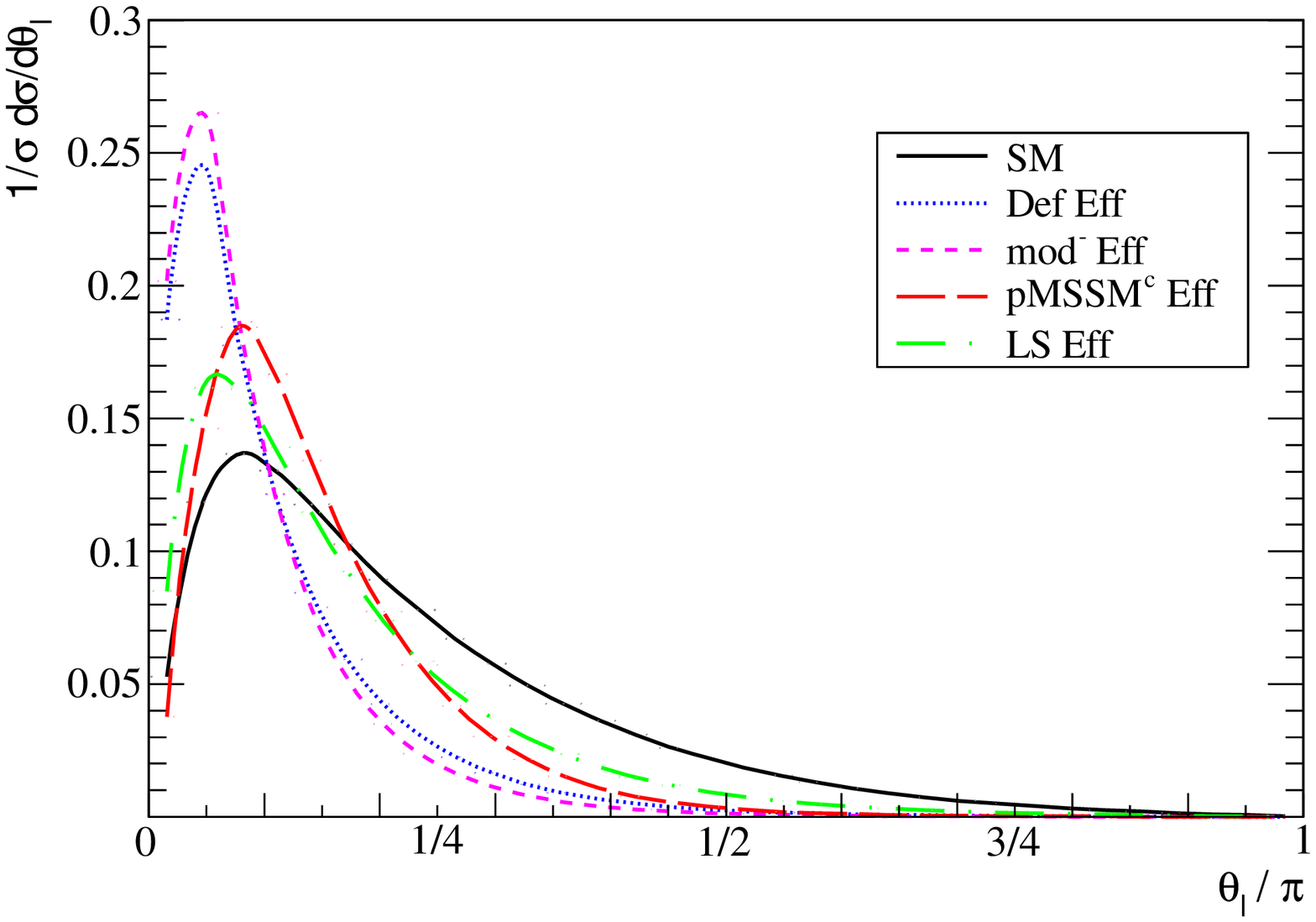}&
\includegraphics[width=6.0cm,angle=0]{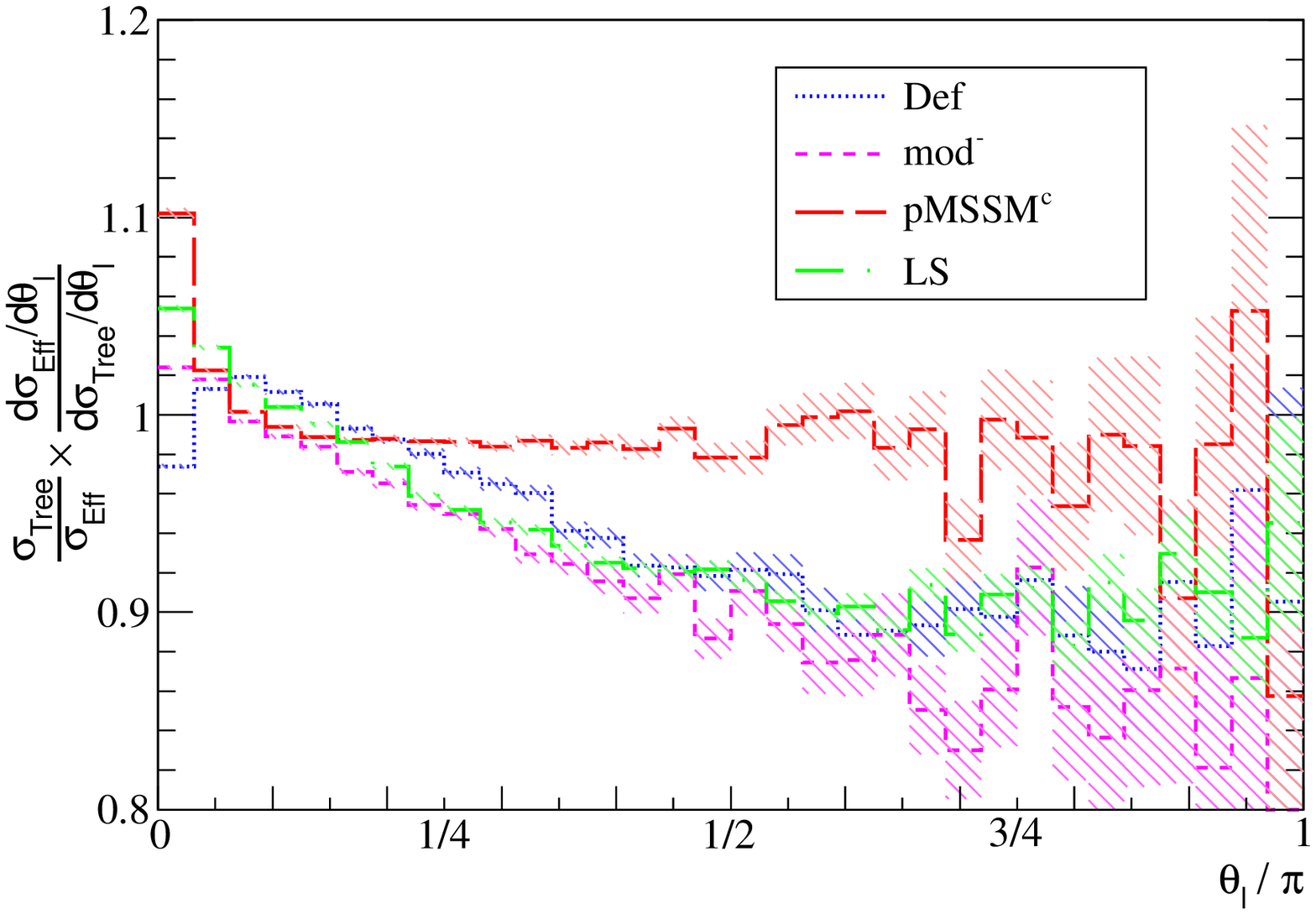}\\
\multicolumn{2}{c}{(b)}\\
\end{tabular}
}
\caption{Normalized cross-section distribution as a function of {\bf (a)} the azimuthal angle $\phi_{l}$
and {\bf (b)} $\theta_{l}$ of the decay lepton for the SM
and all SUSY scenarios studied at $\sqrt{s}=14\TeV$.
The right panels show the factor $dK_\text{SUSY}/dx$, Eq.~\eqref{eq:ksusynew}.
}
\label{fig:leptonAngDist}
\end{figure}

These changes would also be reflected on the final leptons 
angular distribution and hopefully measured through the asymmetries 
defined as in Eqs.~(\ref{eq:asymobserv1}) and (\ref{eq:asymobserv2}).   
Table~\ref{tab:leptonasymmetries} displays the predictions for the 
angular asymmetries, $A_{\phi_{l}}$ and $A_{\theta_{l}}$,
for the SM and the four MSSM scenarios studied here. 
We have checked that our SM results agree with~\cite{Godbole:2010kr}
under the same cut conditions.
In the last column of this table we also include the values of
$\mathcal{P}_{t}$ for the decay $\tilde t_{1} \to t \tilde
\chi_1^0$ with the SUSY parameters as in table~\ref{tab:allMSSM}. 
As already discussed, polarization values different from 0 translate 
into sharply peaked lepton distribution in $\phi_{l}$ -towards 0 
and $2\pi$- and $\theta_{l}$ -towards 0- as can be observed
in the left column of figure~\ref{fig:leptonAngDist} where the 
normalized distribution of both variables 
are presented for the SM and all MSSM parametrizations. As expected, 
when comparing to SM results larger 
asymmetries values for all MSSM scenarios are obtained. 

\begin{figure}[t]
\centering
\resizebox{\textwidth}{!}{
\begin{tabular}{cc}
\includegraphics[width=6.0cm,angle=0]{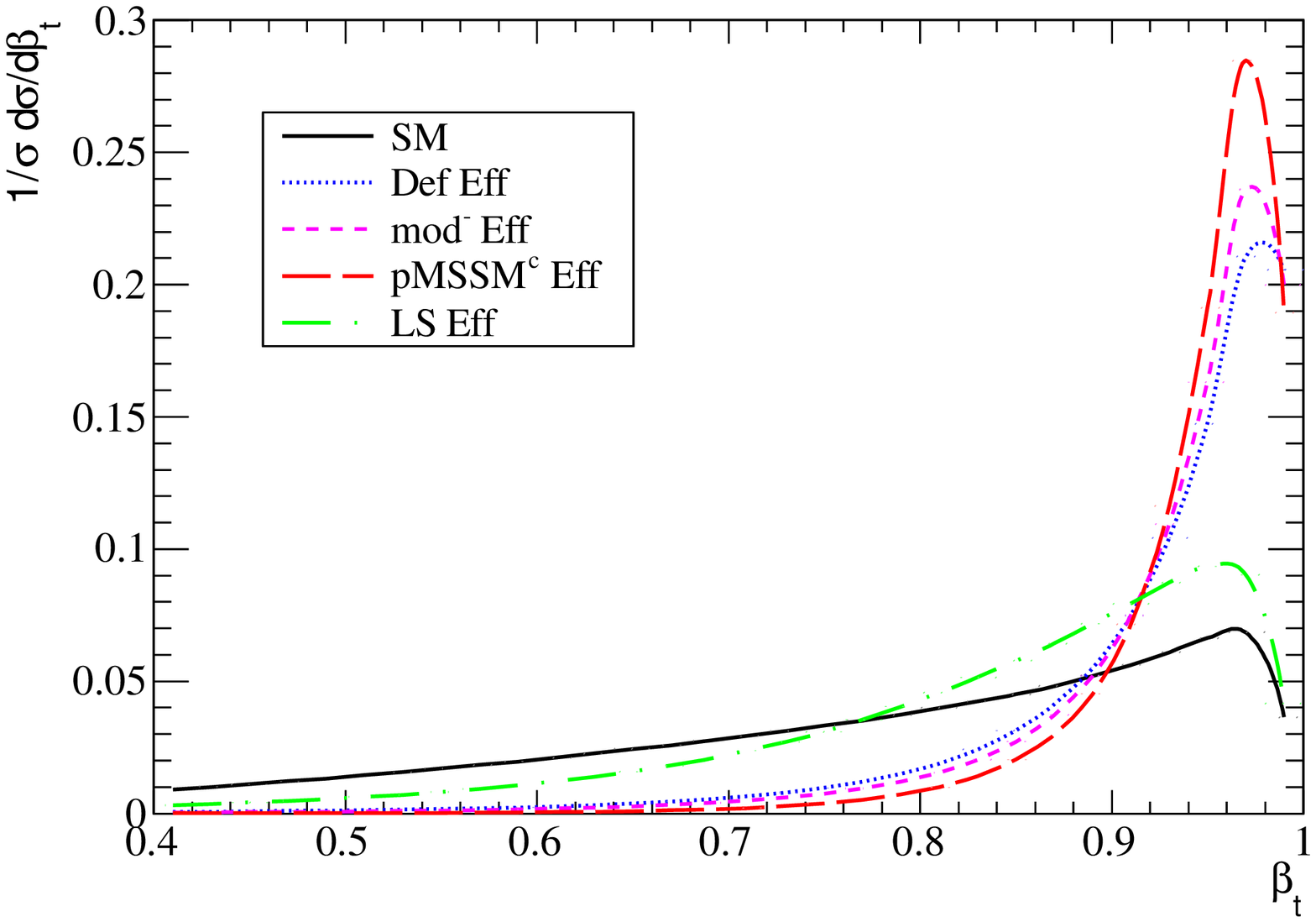}&
\includegraphics[width=6.0cm,angle=0]{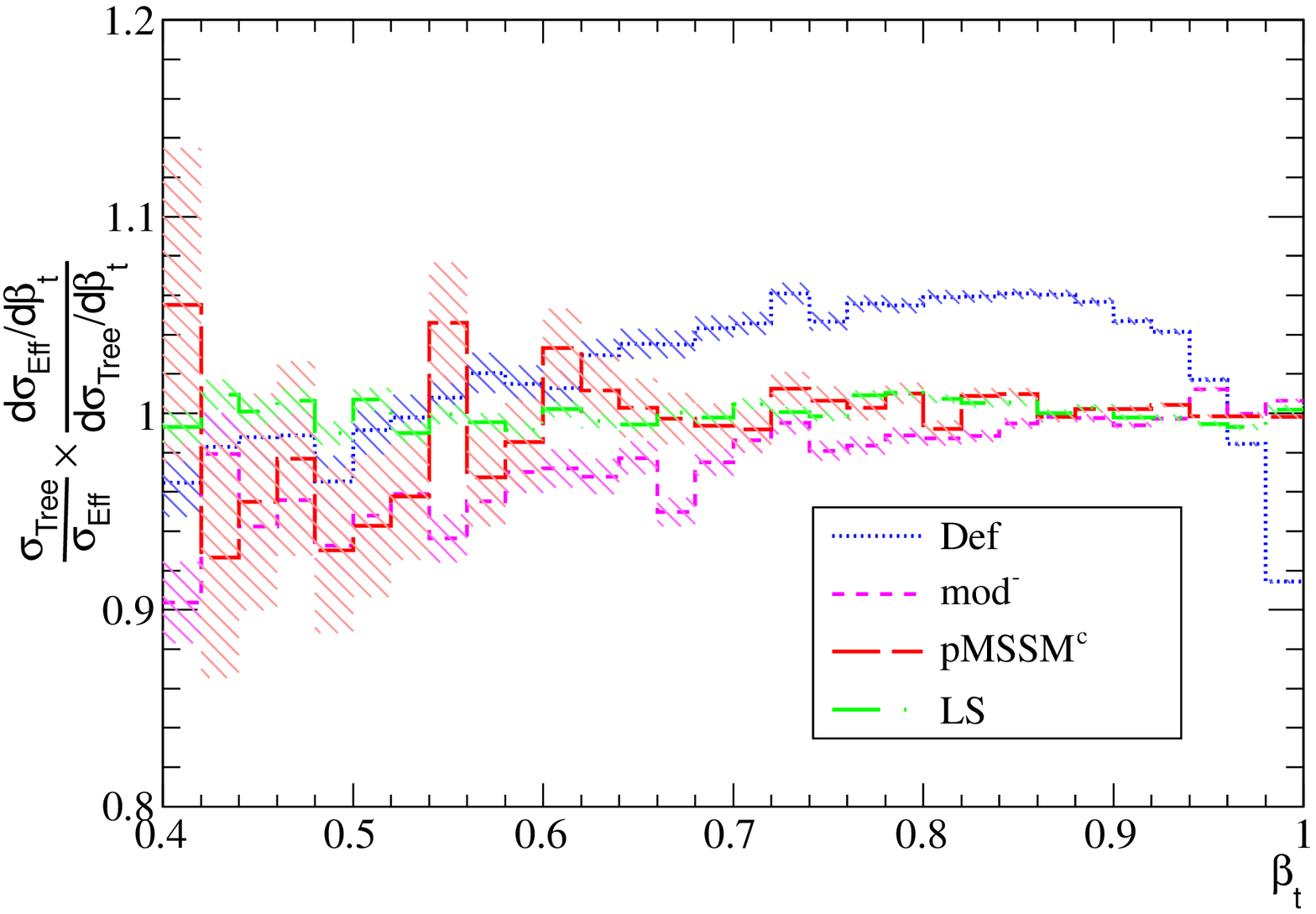}\\
\end{tabular}
}
\caption{As in Figure~\ref{fig:leptonAngDist} but as a function of $\beta_{t}$.}
\label{fig:topboost}
\end{figure}

Figure~\ref{fig:topboost} shows the normalized top-quark boost
distribution, $\beta_{t}$ as in Eq.~(\ref{eq:topboost}), for the SM and all MSSM scenarios.
Note that figures~\ref{fig:leptonAngDist} and~\ref{fig:topboost} only show the result
in the effective approximation. The tree-level results are not presented
here because they are indistinguishable in the plots. 
Top-quark boost distributions are presented for $\beta_{t}>0.4$ 
as it reflects the characteristics of more than 99\% of the top-quark
population for any model or parametrization.
We clearly see how in {\it Def}, $\modmin$ and $\pMSSM$ scenarios a 
large fractions of top quarks ($\gtrsim98\%$) is emitted 
with $\beta_{t}>0.8$, supporting further the sizable values of the
asymmetries obtained for those scenarios. 
The highly boosted top-quark populations smear the possible changes in the asymmetries due to $\mathcal{P}_{t}$
variations. For example, in the {\it{Def}} scenario, a change in $\mathcal{P}_{t}$ of $+130\%$ due to radiative
corrections translates into a tiny $\sim0.4\%$, $\sim1.06\%$ change in $A_{\phi_{l}}$, $A_{\theta_{l}}$ respectively. 
Opposite, the $LS$ scenario shows the less boosted top-quark populations (Figure~\ref{fig:topboost}),
and it has the smallest asymmetries (table~\ref{tab:leptonasymmetries}), 
hence the asymmetries are more sensitive to $\mathcal{P}_{t}$ variations. 
A $\mathcal{P}_{t}$ change of 37\% due to radiative corrections translates 
into a $1.3\%$, $4.5\%$ change in 
 $A_{\phi_{l}}$, $A_{\theta_{l}}$ respectively.
Even though the {\it LS} scenario gives the lowest asymmetries among our 
SUSY scenarios, the predictions in the effective approximation are
a $40\%$ and $88\%$ larger than the SM values for  $A_{\phi_{l}}$ and 
$A_{\theta_{l}}$ respectively.

To asses the effects of the radiative corrections, we define the ratio of the corrected-to-tree-level
distributions,
\begin{equation}
\frac{dK_\text{SUSY}}{dx}= \frac{\sigma_\text{tree}}{\sigma_\text{eff}}\frac{d\sigma_\text{eff}/dx}{d\sigma_\text{tree}/dx}\, ,
\label{eq:ksusynew}
\end{equation}
where $x$ is a distribution variable. Right panels of
figures~\ref{fig:leptonAngDist} and~\ref{fig:topboost}
show the effects of the radiative corrections  on the 
$\phi_{l}$, $\theta_{l}$ and $\beta_{t}$ distributions.
The radiative corrections change slightly the angular distributions 
in the $\phi_{l}\sim\pi$ and $\theta_{l}\sim\pi$ regions, up to a 10\% 
for $\phi_{l}$ and a 20\% for $\theta_{l}$. 
The exception is the \pMSSM\ scenario whose 
$dK_\text{SUSY}/d\phi_{l}$ and $dK_\text{SUSY}/d\theta_{l}$~\eqref{eq:ksusynew} factors
are $\sim1$ in the whole range. As for the $\beta_{t}$ distribution the 
$dK_\text{SUSY}/d\beta_{t}$~\eqref{eq:ksusynew} 
factor is $\sim1$ for all scenarios, except {\it Def}. For {\it Def}
$dK_\text{SUSY}/d\beta_{t}$ is 
$\sim1.05$ for $0.7<\beta_{t}<0.9$ and 0.9 at the largest $\beta_{t}$ bin.
 
\begin{figure}[t]
\centering
\resizebox{\textwidth}{!}{
\begin{tabular}{cc}
\includegraphics[width=6.0cm,angle=0]{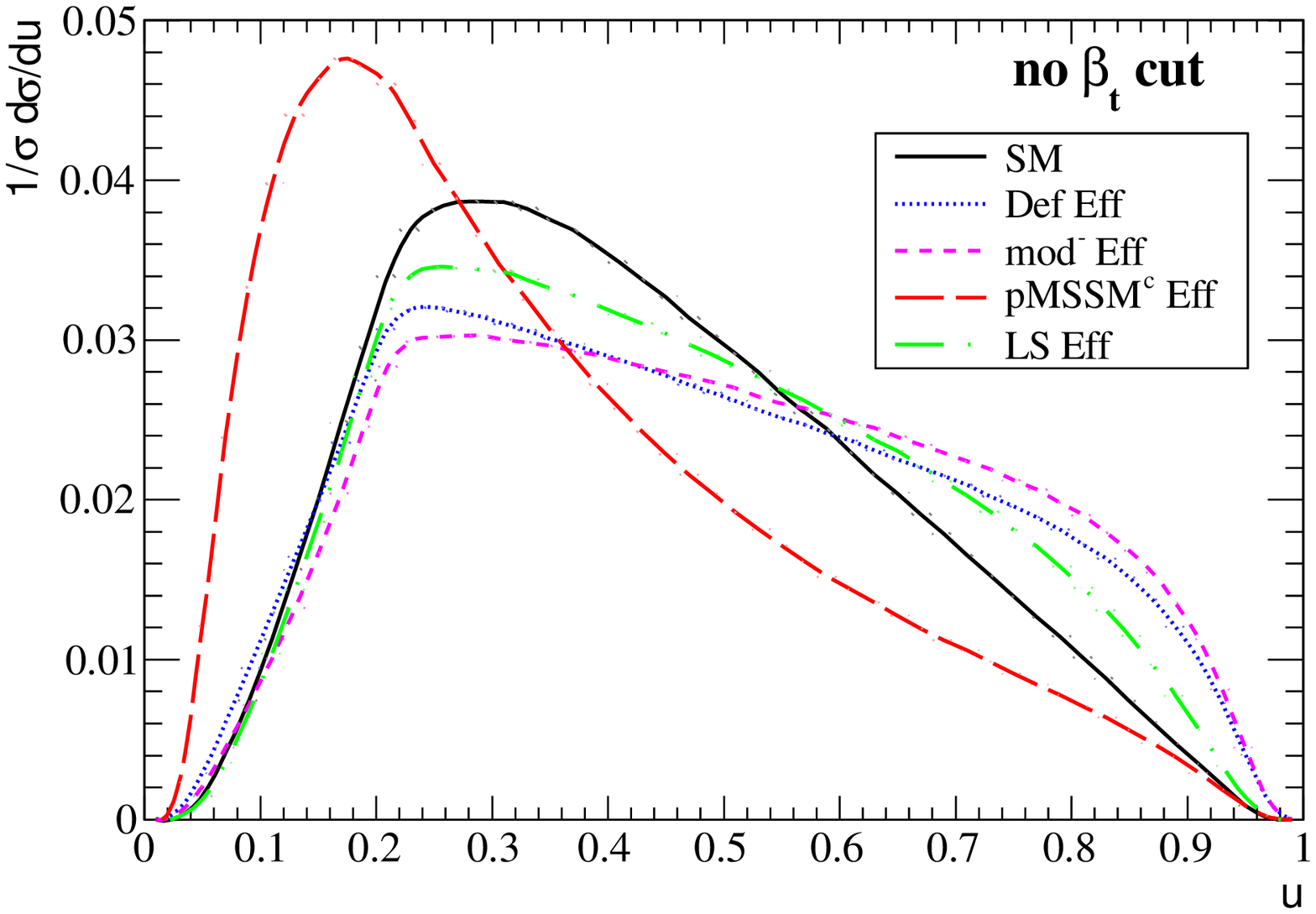}&
\includegraphics[width=6.0cm,angle=0]{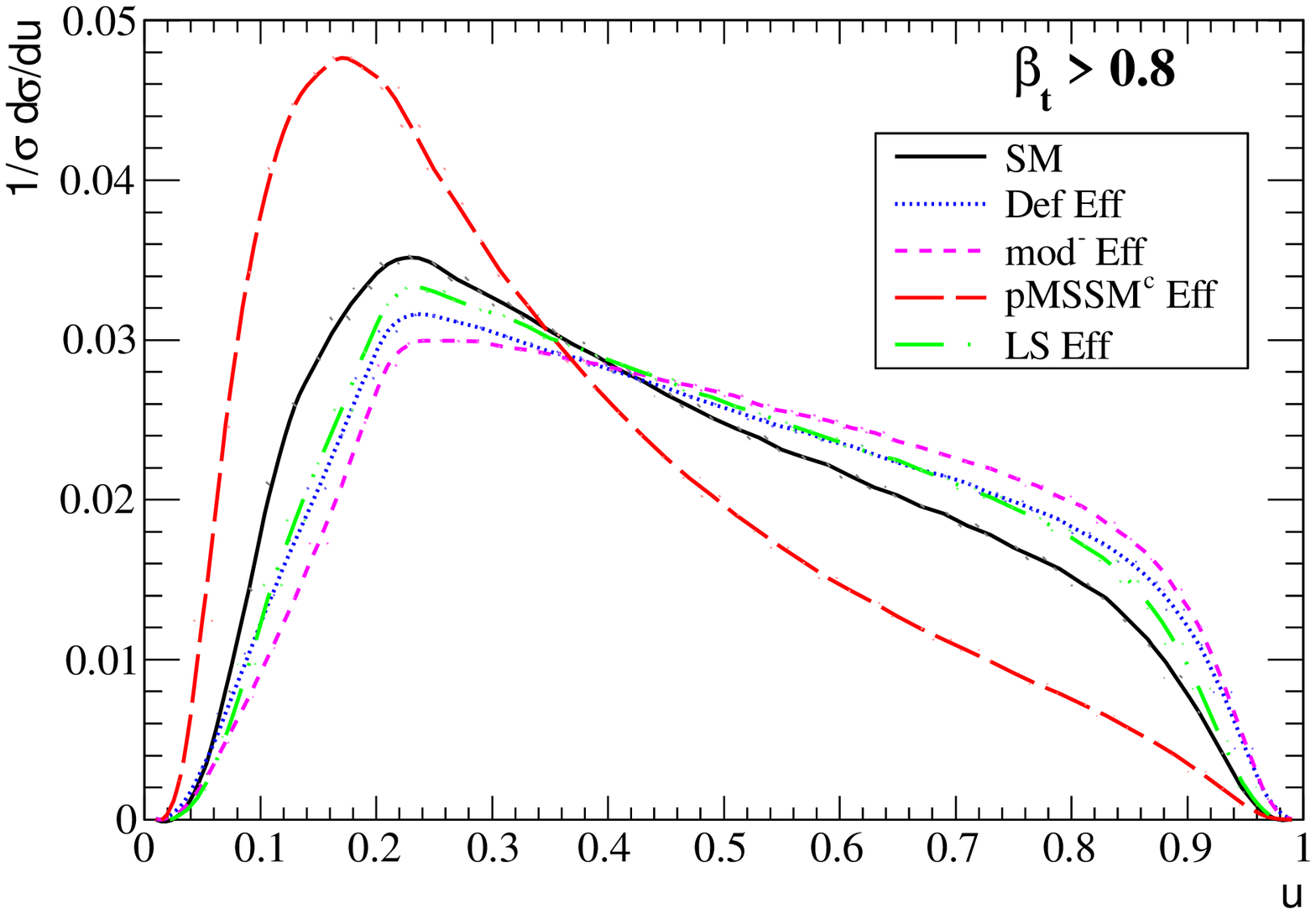}\\
(a)&(b)\\
\end{tabular}
}
\caption{Distribution of the energy ratio $u$ {\bf(a)} without cut on $\beta_{t}$ , 
and {\bf(b)} for $\beta_{t}>0.8$ for the SM and all SUSY scenarios at $\sqrt{s}=14\TeV$.}
\label{fig:ufracallScen}
\end{figure}

\begin{figure}[t]
\centering
\resizebox{\textwidth}{!}{
\begin{tabular}{cc}
\includegraphics[width=6.0cm,angle=0]{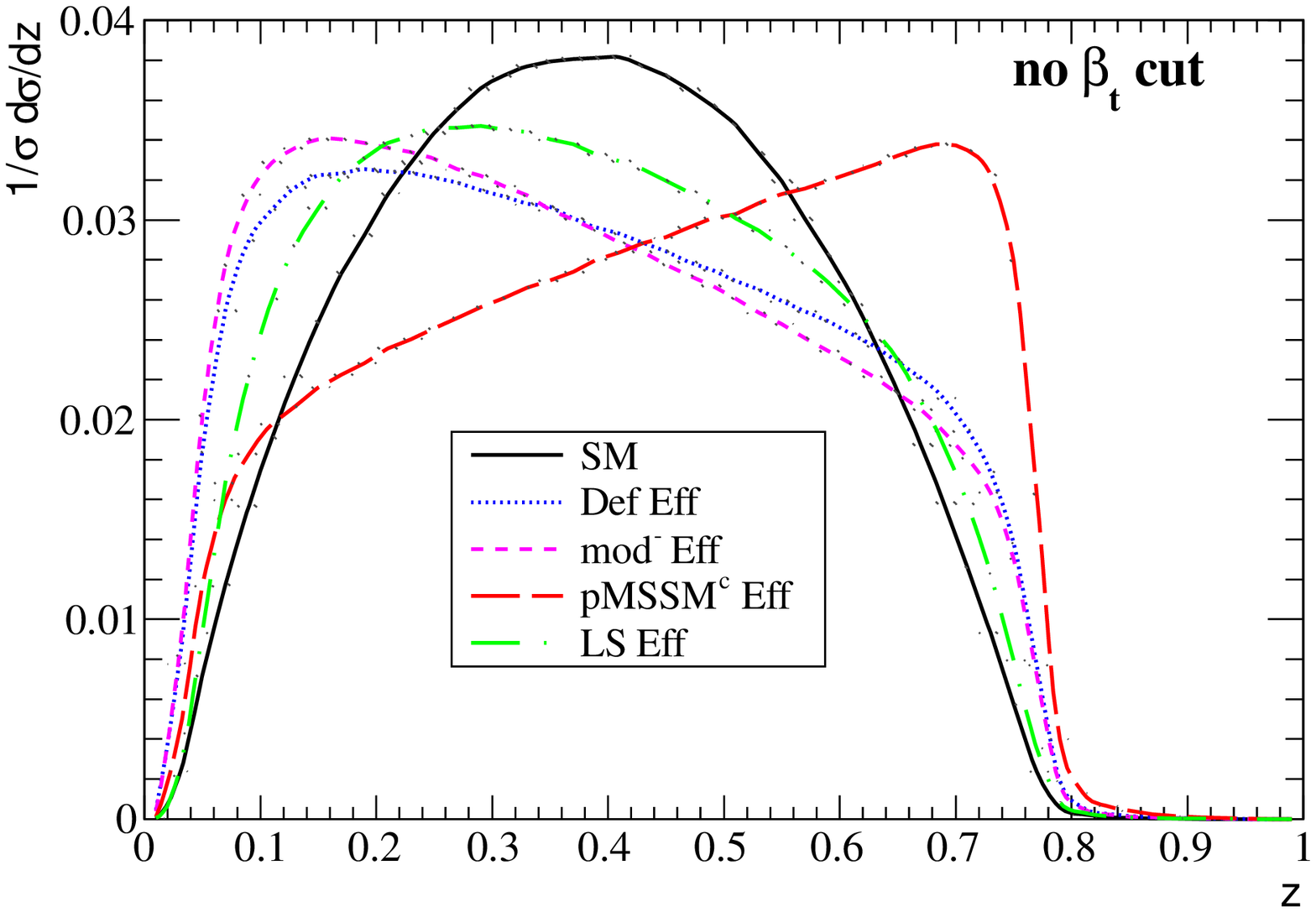}&
\includegraphics[width=6.0cm,angle=0]{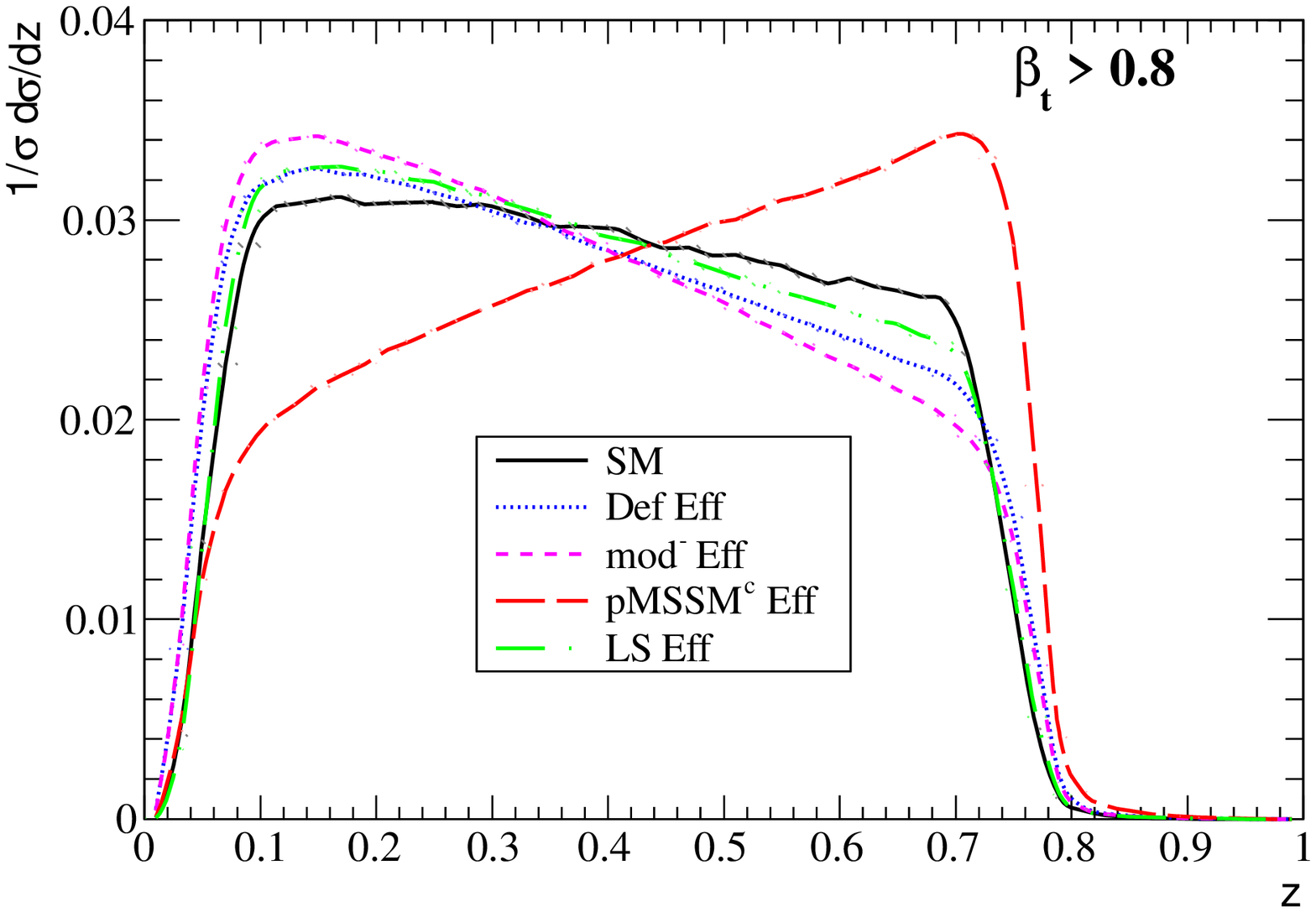}\\
(a)&(b)
\end{tabular}
}
\caption{As in Figure~\ref{fig:ufracallScen} but as a function of $z$.}
\label{fig:zfracallScen}
\end{figure}

As mentioned in~\cite{Belanger:2013gha} and references therein, 
the measurements of lepton angular asymmetries
in the highly boosted top-quark scenarios is a challenge at the LHC. 
Thus, distribution of $u$ and $z$ variables, defined
as in Eqs.~(\ref{eq:zfrac})-(\ref{eq:ufrac}), may serve as a 
better discriminator for new physics scenarios enhancing 
top-quark transverse polarization. 
Figures~\ref{fig:ufracallScen} and~\ref{fig:zfracallScen} show the normalized distribution
of the $u$ and $z$ variables for the SM and the four MSSM scenarios
in the effective description approximation of squark interactions, 
{\bf (a)} without any cut on $\beta_{t}$
and {\bf (b)} requiring a kinematic cut $\beta_{t}>0.8$.
We have checked that larger kinematic cuts do not change the distribution.
In agreement with~\cite{Belanger:2013gha},
the $\pMSSM$ scenario, with the negative polarization,
show the largest difference with respect the 
SM with a shift on the $u$ distribution about 0.1 units toward lower values.
The positively polarized top-quark scenarios
($LS$, $\modmin$ and {\it Def}) differ 
from the SM predictions only slightly in the range $0.7<u<0.95$, a situation 
that worsens when the cut on $\beta_{t}$ is required.
For the $z$ distributions, the positively polarized top-quark scenarios
are hard to distinguish among each other independently
of the $\beta_{t}$ cut, their shape resembles the SM one as 
the $\beta_{t}$-cut is increased. 
Regarding to the radiative corrections, 
the $u$ and $z$ distributions are hardly affected. 
The ratio between the effective and tree-level distributions 
differ on a couple of units of percent in boundary values of the variables: 
$u,z\approx0$ and $u,z\approx1$. This ratio is close to 1 in the rest of
the numerical intervals of these two parameters. 
Therefore, these observables are unaffected by the higher 
order corrections in the quark-squark-chargino/neutralino interactions.

Going back to the discussion about scenarios with more promising signal significances
(see paragraph just above section~\ref{top-asymmetries}), we stress that
the distribution of the energy ratio $z$ does not change significantly,
in the context of the radiative corrections to the quark-squark-chargino/neutralino interactions.

\section{Conclusions}
\label{conclusions}

We have computed and analyzed the top-quark polarizations
and $t\bar t$ charge asymmetries, induced by top-squark pair 
production at the LHC and the subsequent decays $\stopp_{1}\to t\tilde \chi_1^0$. 
The computations have been performed with {\tt MadGraph},
including the effective description of squark interactions with 
charginos and neutralinos for the MSSM case. 
We have considered four different SUSY scenarios, 
as presented in table~\ref{tab:allMSSM}, and we
have focused on the effects of the effective approximation 
of squark interactions in $t\bar t$ charge
asymmetries and top-quark polarization observables as defined in 
section~\ref{observables}. 
We compare the results with the SM expectations, computed using the 
leading NLO QCD corrections as included in {\tt MadGraph} 5.

The SUSY contributions to top-quark charge asymmetries, 
Eqs.~\eqref{eq:AC} and~\eqref{eq:ACymean}, 
are strongly dependent 
on the SUSY scenarios, change significantly with the inclusion of the radiative corrections
and our estimations are of the same order of the SM predictions 
(tables~\ref{tab:ACtt} and~\ref{tab:topasymmetriesymean}).
Using a kinematic cut $Y_{\text{cut}}>1.5$ enhances the SUSY contributions. 
Unfortunately, they are still smaller than current experimental sensitivity. 

For the top-quark polarization studies, 
the radiative corrections to the quark-squark-char\-gino/neutra\-lino vertices increase
the value of the polarization of the top-quarks coming from the squark decay,  
compared to the tree-level prediction (table~\ref{tab:polarizationResume}). 
We have discussed in detail the behavior of the top-quark polarization in all the
SUSY scenarios studied in this work, in which $\tilde{\chi}_1^0$ is a 
pure bino-like neutralino.
The top-quark polarization varies between $-1$ and $1$, 
depending also on the mixing of the
top-squark sector and the mass difference between the top-squark  
and the neutralino. 
The changes in the top-quark polarization induce a change in the 
distributions and asymmetries of the lepton coming from top-quark 
decays (table~\ref{tab:leptonasymmetries}).
Because the top-quarks are highly boosted (Figure~\ref{fig:topboost}), 
the polarization effects are smeared 
in the final distribution. The effects are more visible in scenarios 
with less boosted top-quarks ({\it LS}).

The discrimination power of the $u$ and $z$ variables, 
Eqs.~(\ref{eq:zfrac}) and~(\ref{eq:ufrac}), 
of MSSM scenarios against SM was also tested.
The scenario with negative polarization of the top quarks, $\pMSSM$,
shows the maximal discrimination power against the SM 
independently of the kinematic cut applied on the top-quark boost.
In this analysis we obtain that the inclusion of the higher order 
corrections through the effective description of the squark
interactions does not change strongly the $u$ and $z$ distributions 
obtained at leading order 
(Figures~\ref{fig:ufracallScen} and~\ref{fig:zfracallScen}). 
 
We have found strong effects of the radiative corrections on the top-quark polarization. One would welcome new 
top-quark observables and strategies aimed at analyzing top-quark polarization.

\section*{Acknowledgments}

A.~A. have been supported by the Spanish MICINN grant FPA2012-35453 and a
SANTANDER Scholarship Program for Latinoamerican students. 
E.~A. and S.~P. have been supported by the Spanish DGIID-DGA grant 2013-E24/2 and
the Spanish MICINN grant FPA2012-35453. S.~P. is also supported by DURSI
2009-SGR-502.
The Spanish Consolider-Ingenio 2010 Program CPAN (CSD2007-00042) has also supported this work,
and E.~A. particularly acknowledges this funding.
S.~P. thanks Germ\'an Rodrigo for useful discussions and J.~Guasch for a
careful reading of the manuscript.

\end{document}